\documentclass[twocolumn]{aastex61}
\pdfoutput=1 %for arXiv submission
\usepackage{amsmath,amstext}
\usepackage[T1]{fontenc}
\usepackage{apjfonts} 
\usepackage[figure,figure*]{hypcap}

\shorttitle{SPIRITS\,16tn in NGC~3556}
\shortauthors{Jencson et al.}

\begin{document}

\title{SPIRITS\,16\MakeLowercase{tn} in NGC~3556: A heavily obscured and low-luminosity supernova at 8.8~M\MakeLowercase{pc}}

\author{Jacob E. Jencson}
\altaffiliation{National Science Foundation Graduate Research Fellow}
\affiliation{Cahill Center for Astronomy and Astrophysics, California Institute of Technology, Pasadena, CA 91125, USA}
\author{Mansi M. Kasliwal}
\affiliation{Cahill Center for Astronomy and Astrophysics, California Institute of Technology, Pasadena, CA 91125, USA}
\author{Scott M. Adams}
\affiliation{Cahill Center for Astronomy and Astrophysics, California Institute of Technology, Pasadena, CA 91125, USA}
\author{Howard E. Bond}
\affiliation{Department of Astronomy \& Astrophysics, Pennsylvania State University, University Park, PA 16802, USA}
\affiliation{Space Telescope Science Institute, 3700 San Martin Dr., Baltimore, MD 21218, USA}
\author{Ryan M. Lau}
\affiliation{Jet Propulsion Laboratory, California Institute of Technology, 4800 Oak Grove Drive, Pasadena, CA 91109, USA}
\affiliation{Cahill Center for Astronomy and Astrophysics, California Institute of Technology, Pasadena, CA 91125, USA}
\author{Joel Johansson}
\affiliation{Department of Physics and Astronomy, Division of Astronomy and Space Physics, Uppsala University, Box 516, SE 751 20 Uppsala, Sweden}
\author{Assaf Horesh}
\affiliation{Racah Institute of Physics, The Hebrew University, Jerusalem 91904, Israel}
\author{Kunal P. Mooley}
\affiliation{Department of Physics, Astrophysics, University of Oxford, Denys Wilkinson Building, Oxford OX1 3RH, UK}
\author{Robert Fender}
\affiliation{Department of Physics, Astrophysics, University of Oxford, Denys Wilkinson Building, Oxford OX1 3RH, UK}
\author{Kishalay De}
\affiliation{Cahill Center for Astronomy and Astrophysics, California Institute of Technology, Pasadena, CA 91125, USA}
\author{D\'{o}nal O'Sullivan}
\affiliation{Cahill Center for Astronomy and Astrophysics, California Institute of Technology, Pasadena, CA 91125, USA}
\author{Frank J. Masci}
\affiliation{Caltech/IPAC, Mailcode 100-22, Pasadena, CA 91125, USA}
\author{Ann Marie Cody}
\affiliation{NASA Ames Research Center, Moffet Field, CA 94035}
\author{Nadia Blagorodnova}
\affiliation{Cahill Center for Astronomy and Astrophysics, California Institute of Technology, Pasadena, CA 91125, USA}
\author{Ori D. Fox}
\affiliation{Space Telescope Science Institute, 3700 San Martin Dr., Baltimore, MD 21218, USA}
\author{Robert D. Gehrz}
\affiliation{Minnesota Institute for Astrophysics, School of Physics and Astronomy, 116 Church Street, S. E., University of Minnesota, Minneapolis, MN 55455, USA}
\author{Peter A. Milne}
\affiliation{University of Arizona, Steward Observatory, 933 N. Cherry Avenue, Tucson, AZ 85721, USA}
\author{Daniel A. Perley}
\affiliation{Dark Cosmology Centre, Niels Bohr Institute, University of Copenhagen, Juliane Maries Vej 30, DK-2100 Copenhagen \O, Denmark}
\affiliation{Astrophysics Research Institute, Liverpool John Moores University, IC2, Liverpool Science Park, 146 Brownlow Hill, Liverpool L3 5RF, UK}
\author{Nathan Smith}
\affiliation{University of Arizona, Steward Observatory, 933 N. Cherry Avenue, Tucson, AZ 85721, USA}
\author{Schuyler D. Van Dyk}
\affiliation{Caltech/IPAC, Mailcode 100-22, Pasadena, CA 91125, USA}

\correspondingauthor{Jacob E. Jencson}
\email{jj@astro.caltech.edu}

\begin{abstract}
We present the discovery by the SPitzer InfraRed Intensive Transients Survey (SPIRITS) of a likely supernova (SN) in NGC~3556 (M108) at only 8.8~Mpc, which, despite its proximity, was not detected by optical searches. A luminous infrared (IR) transient at $M_{[4.5]} = -16.7$~mag (Vega), SPIRITS\,16tn is coincident with a dust lane in the inclined, star-forming disk of the host. Using observations in the IR, optical, and radio, we attempt to determine the nature of this event. We estimate $A_{V} \approx 8$--$9$~mag of extinction, placing it among the three most highly obscured IR-discovered SNe to date. The [4.5] light curve declined at a rate of $0.013$~mag~day$^{-1}$, and the $[3.6] - [4.5]$ color grew redder from $0.7$ to $\gtrsim 1.0$~mag by 184.7 days post discovery. Optical/IR spectroscopy shows a red continuum, but no clearly discernible features are evident, preventing a definitive spectroscopic classification. Deep radio observations constrain the radio luminosity of SPIRITS\,16tn to $L_{\nu} \lesssim 10^{24}$~erg~s$^{-1}$~Hz$^{-1}$ between 3--15~GHz, excluding many varieties of radio core-collapse SNe. A type Ia SN is ruled out by the observed red IR color, and lack of features normally attributed to Fe-peak elements in the optical and IR spectra. SPIRITS\,16tn was fainter at [4.5] than typical stripped-envelope SNe by $\approx 1$~mag. Comparison of the spectral energy distribution (SED) to SNe~II suggests SPIRITS\,16tn was both highly obscured, and intrinsically dim, possibly akin to the low-luminosity SN~2005cs. We infer the presence of an IR dust echo powered by an initial peak luminosity of the transient of $5 \times 10^{40}$~erg~s$^{-1} \lesssim L_{\mathrm{peak}} \lesssim 4\times10^{43}$~erg~s$^{-1}$, consistent with the observed range for SNe~II. This discovery illustrates the power of IR surveys to overcome the compounding effects of visible extinction and optically sub-luminous events in completing the inventory of nearby SNe.

\end{abstract}


\section{Introduction} \label{sec:intro}
The discovery and characterization of core-collapse supernovae (CC SNe), bursts of light heralding the explosive deaths of stars with initial mass $\gtrsim 8~M_{\odot}$, has been largely driven in recent years by several large optical time-domain surveys, many specifically dedicated to the identification of transients. While such searches have been hugely successful, now discovering hundreds of SNe every year, a primary limitation is the susceptibility of visible photons to extinction by intervening dust. CC SNe in particular, often associated with the dense and dusty star-forming regions of late-type galaxies, may be subject to significant host extinction.

The measurement of the CCSN rate from optical surveys is an important probe of star formation and the fate of massive stars. However, these measurements only yield lower limits, as some SNe are missed due to obscuration \citep[e.g.,][]{grossan99,maiolino02,cresci07}. In particular, \citet{horiuchi11} claim that half of all supernovae are missing across redshifts from $0 < z < 1$, termed the ``supernova rate problem'' and possibly indicating a large population of hidden or intrinsically dim SNe. \citet{cappellaro15} have challenged this claim, however, finding full agreement between CC SN rates and revised measurements of the cosmic star formation history. Still, \citet{mattila12} find empirically that $\sim 20$\% of SNe locally, growing to $\sim 40$\% by $z=1$, may be missed by optical searches due only to obscuration by dust. The deep, galaxy-targeted $D < 40$~Mpc (DTL40) supernova search recently reported, for example, the discovery of the obscured type II SN DLT16am (SN~2016ija) in the nearby, edge-on galaxy NGC~1532 with $A_V \approx 6$~mag \citep{tartaglia18}. Further confounding the debate, recent studies suggest that CC SNe may even be overproduced in the local 11 Mpc volume \citep{horiuchi13,botticella12,xiao15} compared to H$\alpha$ and UV inferred star formation rates. Any CC SNe missed in the nearest galaxies only increase this tension.

%[Make sure refs are up to date, e.g., and SUNBIRD refs.]

Transient surveys at infrared (IR) wavelengths can overcome the limitations of optical searches introduced by the effects of extinction. A number searches in the near-IR have focused specifically on the dense, highly star-forming, heavily extinguished environments of luminous and ultra-luminous infrared galaxies (LIRGs and ULIRGs), where the SN rates are expected to be high, of order one per year \citep{mattila01}. Such surveys, using seeing-limited imaging \citep[e.g.,][]{mannucci03,miluzio13}, or high-resolution imaging from space or with ground-based adaptive optics to probe the densest nuclear regions of these galaxies \citep[e.g.,][]{cresci07,mattila07,kankare08,kankare12,kool18} have now uncovered 16 CC~SNe in (U)LIRGs. 

The InfraRed Array Camera (IRAC; \citealp{fazio04}) aboard the \textit{Spitzer Space Telescope} (\textit{Spitzer}; \citealp{werner04,gehrz07}), in the 3.6 and 4.5~$\mu$m imaging bands (hereafter [3.6] and [4.5]) where the effects of extinction are minimal, is sensitive to even the most highly obscured events, up to $A_V \approx 100$~mag at 20~Mpc. Since December 2014, the SPitzer InfraRed Intensive Transients Survey (SPIRITS; PIDs 11063, 13053; PI M. Kasliwal, \citealp{kasliwal17}) has been conducting an ongoing monitoring campaign of nearby galaxies ($D\lesssim 20$~Mpc) for transients with \textit{Spitzer}/IRAC at [3.6] and [4.5]. An example of the importance of IR surveys was demonstrated in \citet{jencson17}, where we reported the discovery of two obscured SNe in IC\,2163, SPIRITS\,14buu and SPIRITS\,15C, missed by optical searches despite their proximity to Earth and only moderate level of extinction ($A_V \approx 1.5$--$2.2$~mag).

Beyond the ability to discover CC SNe hidden by dust, mid-IR observations offer important diagnostics of the explosions and their circumburst environments. Mid-IR emission may be produced in SNe as thermal emission from the photosphere of the explosions, and also traces the presence of warm dust in the system. This dust may be newly formed in the ejecta or in the rapidly cooling, post-shock material of the explosion. Alternatively, pre-existing, circumburst dust, possibly formed in the pre-SN stellar wind or an eruptive mass loss event of the progenitor star, may be heated by the luminous SN peak, producing an ``IR echo'' due to light travel time effects \citep[e.g.,][]{bode80,dwek83,mattila08}. The multi-faceted effects of dust, either newly formed or pre-existing, on the mid-IR emission of SNe has been studied in numerous works \citep[see, e.g.,][]{kotak09,fox10,fox11,szalai13}. Most recently, \citet{tinyanont16} presented a systematic study of CC SNe observed in the mid-IR with SPIRITS, finding remarkable diversity in the growing sample of well characterized events at these wavelengths.

%[Add paragraph on Radio]

Here, we report the discovery of SPIRITS\,16tn, a likely highly obscured CC SN at only 8.8~Mpc in the nearby spiral galaxy NGC~3556 (M108). In Section~\ref{sec:discovery}, we describe the discovery and follow-up observations of this event using both space and ground-based facilities in the optical, IR, and radio. In Section~\ref{sec:analysis}, we describe our analysis of the data, including constraints on the progenitor luminosity from archival, pre-explosion \textit{Hubble Space Telescope} (\textit{HST}) imaging (Section~\ref{sec:progenitor}), analysis of the light curves and color evolution (Section~\ref{sec:lcs}), evolution of the spectral energy distribution (SED) and constraints on the extinction and dust emission (Section~\ref{sec:sed}). In Section~\ref{sec:discussion}, we discuss of the overall properties of SPIRITS\,16tn and our interpretation of the observations in the context of well-studied SNe and other types of luminous IR transients. We present our conclusions in Section~\ref{sec:conclusions}. 

\section{SPIRITS discovery and follow-up observations} \label{sec:discovery}

\subsection{\textit{Spitzer}/IRAC Discovery in NGC~3556}\label{sec:Spitzer}
\begin{figure*}
\begin{minipage}[hpb]{180mm}
\centering
\includegraphics[width=0.9\linewidth]{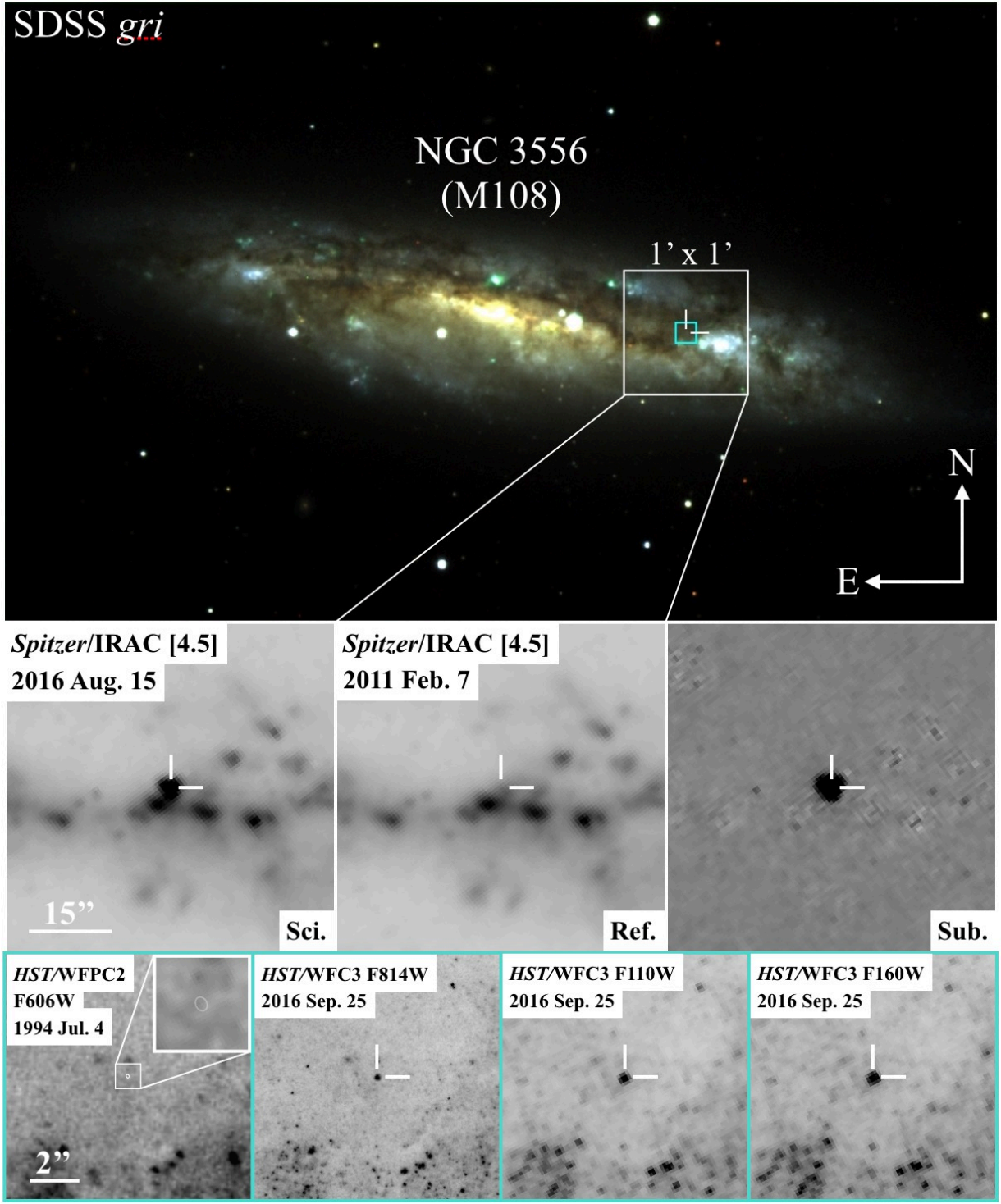}
\caption{\label{fig:images}
The top panel shows the color-composite SDSS imaging of NGC~3556 (M108) in three filters ($g$ in blue, $r$ in green, and $i$ in red). The location of SPIRITS\,16tn in a dust lane of NGC~3556 is indicated by the white crosshairs. The middle row shows the $1~\mathrm{arcmin} \times 1~\mathrm{arcmin}$ region indicated by the white zoom-in box in the top panel. From left to right, we show the \textit{Spitzer}/IRAC [4.5] discovery science frame of SPIRITS\,16tn from 2016 August 15, the archival reference image from 2011 February 7, and the $\mathrm{science} - \mathrm{reference}$ subtraction image, clearly showing the new transient source. In the bottom row, we show the $10~\mathrm{arcsec} \times 10~\mathrm{arcsec}$ region indicated by the cyan box in the top panel. In the leftmost panel, we show the archival \textit{HST}/WFPC2 F606W image. The white ellipse (shown more clearly in the $1~\mathrm{arcsec} \times 1~\mathrm{arcsec}$ zoom-in in the upper-right corner of this panel) indicates the 10-$\sigma$ uncertainty on the position of SPIRITS\,16tn. In the three rightmost panels of the bottom row, we show the post-discovery $HST$/WFC3 F814W, F110W, and F160W imaging of SPIRITS\,16tn. 
}
\end{minipage}
\end{figure*}

During the ongoing monitoring campaign of nearby galaxies with SPIRITS, we observed the star-forming galaxy NGC~3556 with \textit{Spitzer}/IRAC at [3.6] and [4.5] at 8 epochs between UT 2014 January 18.4 and 2016 August 15.0. Image subtraction was performed using archival images from 2011 February 7.6 as references (observed as part of the \textit{Spitzer} Survey of Stellar Structure in Galaxies, S$^4$; PID 61065; PI K. Sheth; \citealp{sheth10}). For details on our image subtraction pipeline see \citet{kasliwal17}. A new transient source, designated SPIRITS\,16tn, was detected in both the [3.6] and [4.5] images on 2016 August 15.0 (MJD = 57615.0; \citealp{jencson16}). Throughout this paper, we refer to the phase as the number of days since the earliest detection of SPIRITS\,16tn on this date. We detect no significant variability at the location of SPIRITS\,16tn in any of the prior Spitzer/IRAC images compared to the reference frame. We show the [4.5] discovery images in the center row of Figure~\ref{fig:images}, along with mosaicked $gri$ imaging of the field from the Sloan Digital Sky Survey Date Release 12 (SDSS-DR12; \citealp{eisenstein11,alam15}) in the top panel showing the location of SPIRITS\,16tn in a dust lane in the disk of NGC~3556. This galaxy was also the host of the probable type II SN~1969B \citep{ciatti71}.

\subsubsection{Host distance and Galactic extinction}
SPIRITS\,16tn was discovered at a right ascension and declination of $11^{\mathrm{h}}11^{\mathrm{m}}20\fs40, +55\degr40\arcmin17\farcs3$ (J2000). Located $89.9$~arcsec from the center of the star-forming galaxy NGC~3556, the position of SPIRITS\,16tn is coincident with a dust lane in the disk.

NED\footnote{The NASA/IPAC Extragalactic Database (NED) is operated by the Jet Propulsion Laboratory, California Institute of Technology, under contract with the National Aeronautics and Space Administration.} lists 17 individual distance estimates to NGC~3556 with a median value of $\mu = 29.71$~mag and standard deviation of $0.67$~mag. Throughout this work, we adopt the most recent value from \citet{sorce14} of $\mu=29.72 \pm 0.41$~mag ($D \approx 8.8$~Mpc). This estimate is based on the mid-IR Tully-Fisher relation using the [3.6] micron flux with color and selection bias corrections. The redshift of NGC~3556 is $z = 0.002332$ ($v = 699$~km~s$^{-1}$; \citealp{shostak75}). 

We assume Galactic extinction along the line of sight to NGC~3556 of $A_V = 0.046$~mag from the \citet{schlafly11} recalibration of the \citet{schlegel98} IR-based dust map assuming a \citet{fitzpatrick99} extinction law with $R_V = 3.1$. Furthermore, for all other considerations of the possible extinction to SPIRITS\,16tn throughout this work, including any foreground host extinction, we assume the \citet{fitzpatrick99} Milky Way extinction curve with $R_V = 3.1$ unless otherwise noted. 

\subsection{Follow-up Imaging}\label{sec:imaging}
In this section, we describe our space and ground-based imaging follow-up efforts to characterize SPIRITS\,16tn. 

\subsubsection{Space-based}\label{sec:space_imaging}
Since its discovery, we continued to monitor SPIRITS\,16tn with \textit{Spitzer}/IRAC at [3.6] and [4.5] as part of the SPIRITS program. Image subtraction was performed on all subsequent epochs, as described in Section~\ref{sec:Spitzer}. Photometry was performed on the reference subtracted images using a 4-mosaicked-pixel (2.4~arcsec) aperture and a background annulus from 4--12 pixels (2.4--7.2~arcsec). The extracted flux was multiplied by aperture corrections of 1.215 and 1.233 for [3.6] and [4.5], respectively, as described in the IRAC instrument handbook\footnote{\url{http://irsa.ipac.caltech.edu/data/SPITZER/docs/irac/iracinstrumenthandbook/}}. Fluxes then were converted to Vega system magnitudes using the handbook-defined zero magnitude fluxes for each IRAC channel. At discovery, our photometry gives $[4.5] = 13.04\pm0.05$~mag ($M_{[4.5]} = -16.7$~mag; $\lambda L_{\lambda} = 1.3\times10^{7}~L_{\odot}$).

We triggered observations with the Ultra-Violet/Optical Telescope (UVOT; \citealp{roming05}) on board the \textit{Neil Gehrels Swift Observatory} (\textit{Swift}; \citealp{gehrels04,nousek04}) on 2016 August 29.1. No source was detected in the $U$, $B$, and $V$-band images with integration times of 540, 580, and 540~s, respectively \citep{adams16b}. We derived 5-$\sigma$ limiting magnitudes of $V > 19.9$, $B > 20.2$, and $U > 19.9$~mag. The extreme $V - [4.5] \gtrsim 6.9$~mag color indicates SPIRITS\,16tn is likely highly obscured.

\begin{figure}
\centering
\includegraphics[width=\linewidth]{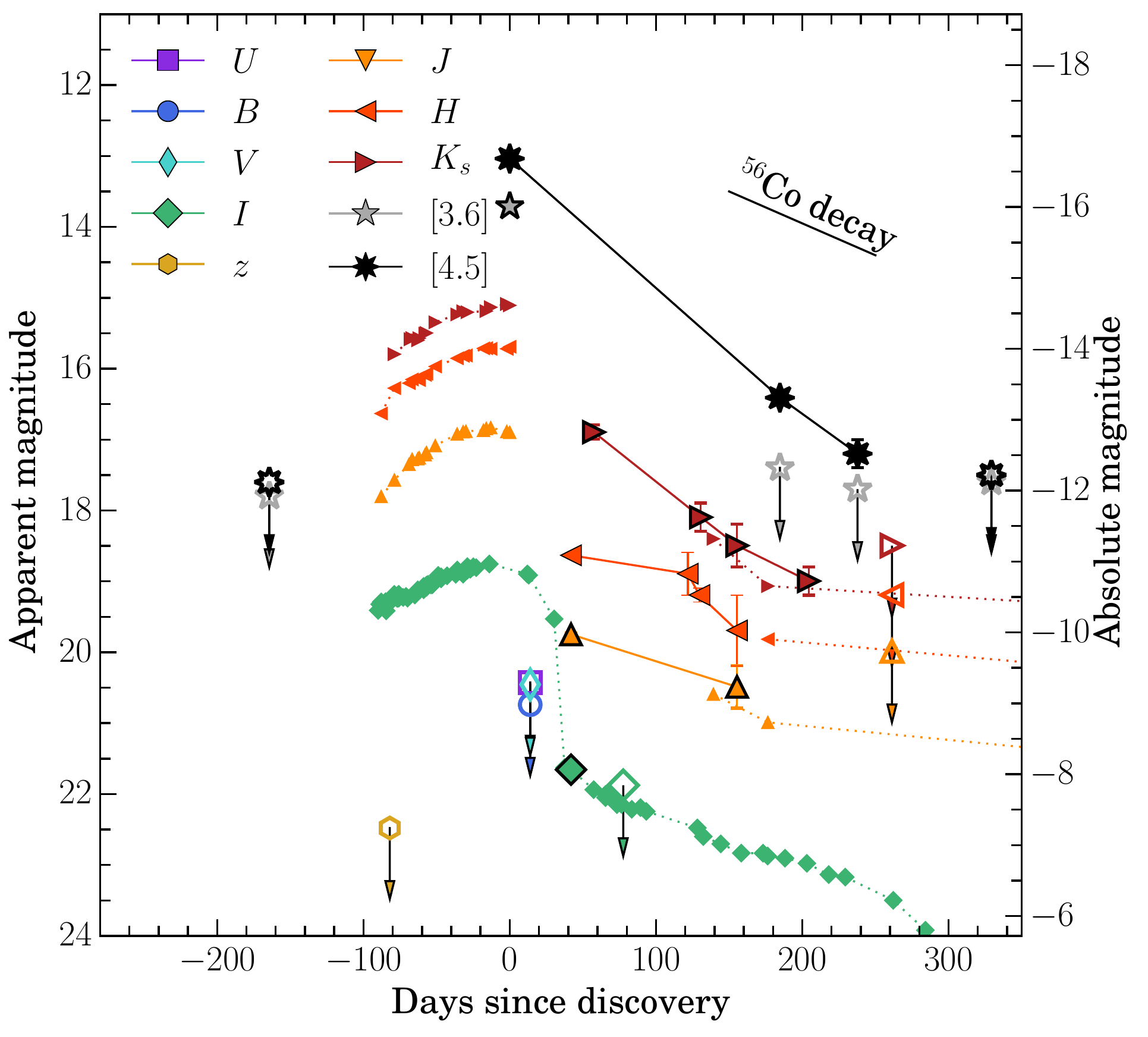}
\caption{\label{fig:lcs}
Multi-band light curves of SPIRITS\,16tn, corrected for Galactic extinction only, are shown as the large, filled symbols, with $5\sigma$ upper limits from non-detections indicated by unfilled symbols with downward arrows. Small symbols are the corresponding light curves of the low-luminosity SN~2005cs from \citep{pastorello06,pastorello09}, shifted to the distance of NGC~3556 plus an additional $\Delta m = 0.7$~mag, and reddened by $E(B-V) = 2.5$~mag. The black, solid line indicates the expected decline rate for a light curve powered by the radioactive decay of $^{56}$Co \citep[see, e.g.,][]{gehrz88,gehrz90}.} 
\end{figure}

We observed SPIRITS\,16tn on 2016 September 25, $t = 42$~days, with the Wide Field Camera 3 (WFC3) on the \textit{Hubble Space Telescope} (\textit{HST}) in the UVIS channel with the F814W filter, and the IR channel with the F110W and F160W filters. These observations were part of our Cycle 23 Target of Opportunity program to observe SPIRITS transients (GO-14258, PI: H. Bond). All three images are shown in the bottom row of Figure~\ref{fig:images}. The photometry and limits from our space-based follow-up effort are listed in Table~\ref{table:phot} and shown in Figure~\ref{fig:lcs}.

\clearpage
\startlongtable
\begin{deluxetable*}{lcccccc}
\tablecaption{Photometry of SPIRITS\,16tn \label{table:phot}}
\tablehead{\colhead{UT Date} & \colhead{MJD} & \colhead{Phase\tablenotemark{a}} & \colhead{Tel./Inst.} & \colhead{Band} & \colhead{Apparent Magnitude\tablenotemark{b,c}} & \colhead{Absolute Magnitude\tablenotemark{c,d}} \\ 
\colhead{} & \colhead{} & \colhead{(days)} & \colhead{} & \colhead{} & \colhead{(mag)} & \colhead{(mag)} } 
\startdata
1999 Jul 04 & 51363 & -6252 & \textit{HST}/WFPC2 & F606W & $>24.5$ & $>-5.2$ \\
2016 Mar 03.6 & 57450.6 & -164.4 & \textit{Spitzer}/IRAC & $[3.6]$ & $>17.8$ & $>-11.9$ \\
2016 Mar 03.6 & 57450.6 & -164.4 & \textit{Spitzer}/IRAC & $[4.5]$ & $>17.6$ & $>-12.1$ \\
2016 May 25.0 & 57533.0 & -82.0 & KPNO-4m & $z$ & $>22.5$ & $>-7.2$ \\
2016 Aug 15.0 & 57615.0 & 0.0 & \textit{Spitzer}/IRAC & $[3.6]$ & $13.71$ $(0.05)$ & $-16.0$ \\
2016 Aug 15.0 & 57615.0 & 0.0 & \textit{Spitzer}/IRAC & $[4.5]$ & $13.04$ $(0.05)$ & $-16.7$ \\
2016 Aug 29.0 & 57629.0 & 14.0 & \textit{Swift}/UVOT & $U$ & $>20.5$ & $>-9.3$ \\
2016 Aug 29.0 & 57629.0 & 14.0 & \textit{Swift}/UVOT & $B$ & $>20.8$ & $>-9.0$ \\
2016 Aug 29.0 & 57629.0 & 14.0 & \textit{Swift}/UVOT & $V$ & $>20.5$ & $>-9.2$ \\
2016 Sep 25.9 & 57656.9 & 41.9 & \textit{HST}/WFC3 & F814W & $21.68$ $(0.03)$ & $-8.0$ \\
2016 Sep 25.9 & 57656.9 & 41.9 & \textit{HST}/WFC3 & F110W & $19.76$ $(0.02)$ & $-10.0$ \\
2016 Sep 25.9 & 57656.9 & 41.9 & \textit{HST}/WFC3 & F160W & $18.64$ $(0.02)$ & $-11.1$ \\
2016 Oct 11.5 & 57672.5 & 57.5 & P200/WIRC & $K_s$ & $16.9$ $(0.1)$ & $-12.8$ \\
2016 Oct 13.5 & 57674.5 & 59.5 & P60/SEDM & $i$ & $>19.4$ & $>-10.3$ \\
2016 Oct 31.6 & 57692.6 & 77.6 & Keck/LRIS & $g$ & $>22.4$ & $>-7.4$ \\
2016 Oct 31.6 & 57692.6 & 77.6 & Keck/LRIS & $I$ & $>21.9$ & $>-7.8$ \\
2016 Nov 09.4 & 57701.4 & 86.4 & 1.5m/RATIR & $r$ & $>19.7$ & $>-10.0$ \\
2016 Nov 09.4 & 57701.4 & 86.4 & 1.5m/RATIR & $i$ & $>19.6$ & $>-10.1$ \\
2016 Nov 09.4 & 57701.4 & 86.4 & 1.5m/RATIR & $z$ & $>19.7$ & $>-10.0$ \\
2016 Nov 09.4 & 57701.4 & 86.4 & 1.5m/RATIR & $Y$ & $>18.6$ & $>-11.1$ \\
2016 Nov 09.4 & 57701.4 & 86.4 & 1.5m/RATIR & $J$ & $>18.0$ & $>-11.7$ \\
2016 Nov 09.4 & 57701.4 & 86.4 & 1.5m/RATIR & $H$ & $>17.5$ & $>-12.2$ \\
2016 Nov 10.4 & 57702.4 & 87.4 & 1.5m/RATIR & $r$ & $>19.9$ & $>-9.8$ \\
2016 Nov 10.4 & 57702.4 & 87.4 & 1.5m/RATIR & $i$ & $>19.8$ & $>-9.9$ \\
2016 Nov 10.4 & 57702.4 & 87.4 & 1.5m/RATIR & $z$ & $>19.7$ & $>-10.0$ \\
2016 Nov 10.4 & 57702.4 & 87.4 & 1.5m/RATIR & $Y$ & $>18.8$ & $>-10.9$ \\
2016 Nov 10.4 & 57702.4 & 87.4 & 1.5m/RATIR & $J$ & $>18.2$ & $>-11.5$ \\
2016 Nov 10.4 & 57702.4 & 87.4 & 1.5m/RATIR & $H$ & $>17.6$ & $>-12.1$ \\
2016 Dec 14.7 & 57736.7 & 121.7 & UKIRT & $H$ & $18.9$ $(0.3)$ & $-10.8$ \\
2016 Dec 22.7 & 57744.7 & 129.7 & UKIRT & $H$ & $19.2$ $(0.1)$ & $-10.5$ \\
2016 Dec 23.6 & 57745.6 & 130.6 & UKIRT & $K$ & $18.1$ $(0.2)$ & $-11.6$ \\
2017 Jan 17.4 & 57770.4 & 155.4 & P200/WIRC & $J$ & $20.5$ $(0.3)$ & $-9.2$ \\
2017 Jan 17.4 & 57770.4 & 155.4 & P200/WIRC & $H$ & $19.7$ $(0.5)$ & $-10.0$ \\
2017 Jan 17.4 & 57770.4 & 155.4 & P200/WIRC & $K_s$ & $18.5$ $(0.3)$ & $-11.2$ \\
2017 Feb 15.7 & 57799.7 & 184.7 & \textit{Spitzer}/IRAC & $[3.6]$ & $>17.4$ & $>-12.3$ \\
2017 Feb 15.7 & 57799.7 & 184.7 & \textit{Spitzer}/IRAC & $[4.5]$ & $16.41$ $(0.07)$ & $-13.3$ \\
2017 Mar 07.4 & 57819.4 & 204.4 & P200/WIRC & $K_s$ & $19.0$ $(0.2)$ & $-10.7$ \\
2017 Apr 09.8 & 57852.8 & 237.8 & \textit{Spitzer}/IRAC & $[3.6]$ & $>17.7$ & $>-12.0$ \\
2017 Apr 09.8 & 57852.8 & 237.8 & \textit{Spitzer}/IRAC & $[4.5]$ & $17.2$ $(0.2)$ & $-12.5$ \\
2017 May 03.3 & 57876.3 & 261.3 & P200/WIRC & $J$ & $>20.0$ & $>-9.7$ \\
2017 May 03.3 & 57876.3 & 261.3 & P200/WIRC & $H$ & $>19.2$ & $>-10.5$ \\
2017 May 03.3 & 57876.3 & 261.3 & P200/WIRC & $K_s$ & $>18.5$ & $>-11.2$ \\
2017 Jul 10.3 & 57944.3 & 329.3 & \textit{Spitzer}/IRAC & $[3.6]$ & $>17.6$ & $>-12.1$ \\
2017 Jul 10.3 & 57944.3 & 329.3 & \textit{Spitzer}/IRAC & $[4.5]$ & $>17.5$ & $>-12.2$ \\
\enddata
\tablenotetext{a}{Phase is number of days since the earliest detection of this event on 2016 August 15.0 ($\mathrm{MJD} = 57615.0$).}
\tablenotetext{b}{Vega magnitudes, except for the $griz$-bands which are AB magnitudes on the SDSS system. 1-$\sigma$ uncertainties are given in parentheses.}
\tablenotetext{c}{5-$\sigma$ limiting magnitudes are given for non-detections.}
\tablenotetext{d}{Absolute magnitudes corrected for Galactic extinction for NGC~3556 from NED.}
\end{deluxetable*}

\subsubsection{Ground-based}\label{sec:ground_imaging}
At the time of its discovery, SPIRITS\,16tn was inaccessible for ground-based observing except at high latitudes. We began ground-based follow-up of SPIRITS\,16tn from Palomar Observatory in 2016 October, approximately 2 months after discovery. We obtained near-IR images of SPIRITS\,16tn in $JHK_s$ with the Wide Field Infrared Camera (WIRC; \citealp{wilson03}) on the Palomar 200-in. Hale Telescope (P200) at several epochs, employing large dithers approximately every minute to allow for accurate subtraction of the bright near-IR sky background. Flat-fielding, background subtraction, astrometric alignment, and final stacking of images in each filter were performed using a custom pipeline. 

Additional near-IR $H$ and $K_s$-band imaging was obtained with the Wide Field Camera (WFCAM; \citealp{casali07}) on the United Kingdom Infrared Telescope (UKIRT) at Mauna Kea Observatory. We obtained simultaneous optical/near-IR $rizYJH$ with the Reionization and Transients InfraRed camera (RATIR; \citealp{butler12}) on the 1.5~m Johnson Telescope at the Mexican Observatorio Astronomico Nacional on the Sierra San Pedro Martir in Baja California, Mexico \citep{watson12}. 

We obtained one epoch of optical ($i$-band) imaging with the Spectral Energy Distribution Machine (SEDM; Blagorodnova et al., in prep) on the fully-automated Palomar 60-in. Telescope (P60; \citealp{cenko06}), and an epoch of $g$ and $I$~band imaging with the Low-Resolution Imaging Spectrometer (LRIS; \citealp{oke95}) on the Keck I Telescope on Mauna Kea. 

Photometry was performed by simultaneously fitting the point-spread function of the transient, measured using field stars, and background, modeled using low-order polynomials. The photometric zero point in each image was obtained by performing photometry on stars of known magnitude in the field. For the near-IR $JHK$ images, we selected 10 bright, isolated 2MASS stars, and for $Y$-band we adopt the conversion from 2MASS used for WFCAM/UKIRT from \citet{hodgkin09}. For optical images, we used 12 SDSS stars, adopting the conversions of \citet{jordi06} to convert from the Sloan $ugriz$ system to $UBVRI$ magnitudes where necessary.

We examined the location of SPIRITS\,16tn in a deep $z$-band image of NGC~3556 from 2017 May 25.0, taken with the CCD Mosaic imager on the 4-m Mayall Telescope at Kitt Peak National Observatory (KPNO) as part of the Mayall z-band Legacy Survey (MzLS). We derive a limit on the flux from the transient of $z > 22.5$~mag, providing our most stringent constraint on the explosion date of SPIRITS\,16tn at 82.0 days before the first detection.

We list all of our photometry of SPIRITS\,16tn in Table~\ref{table:phot}. For non-detections we list 5-$\sigma$ upper limits, where we estimated $\sigma$ as the standard deviation of the pixel values near the transient position to account for uncertain variations in the background flux from the host galaxy. Our light curves of SPIRITS\,16tn are shown in Figure~\ref{fig:lcs}. 

\subsection{Spectroscopy}
\begin{figure}
\centering
\includegraphics[width=\linewidth]{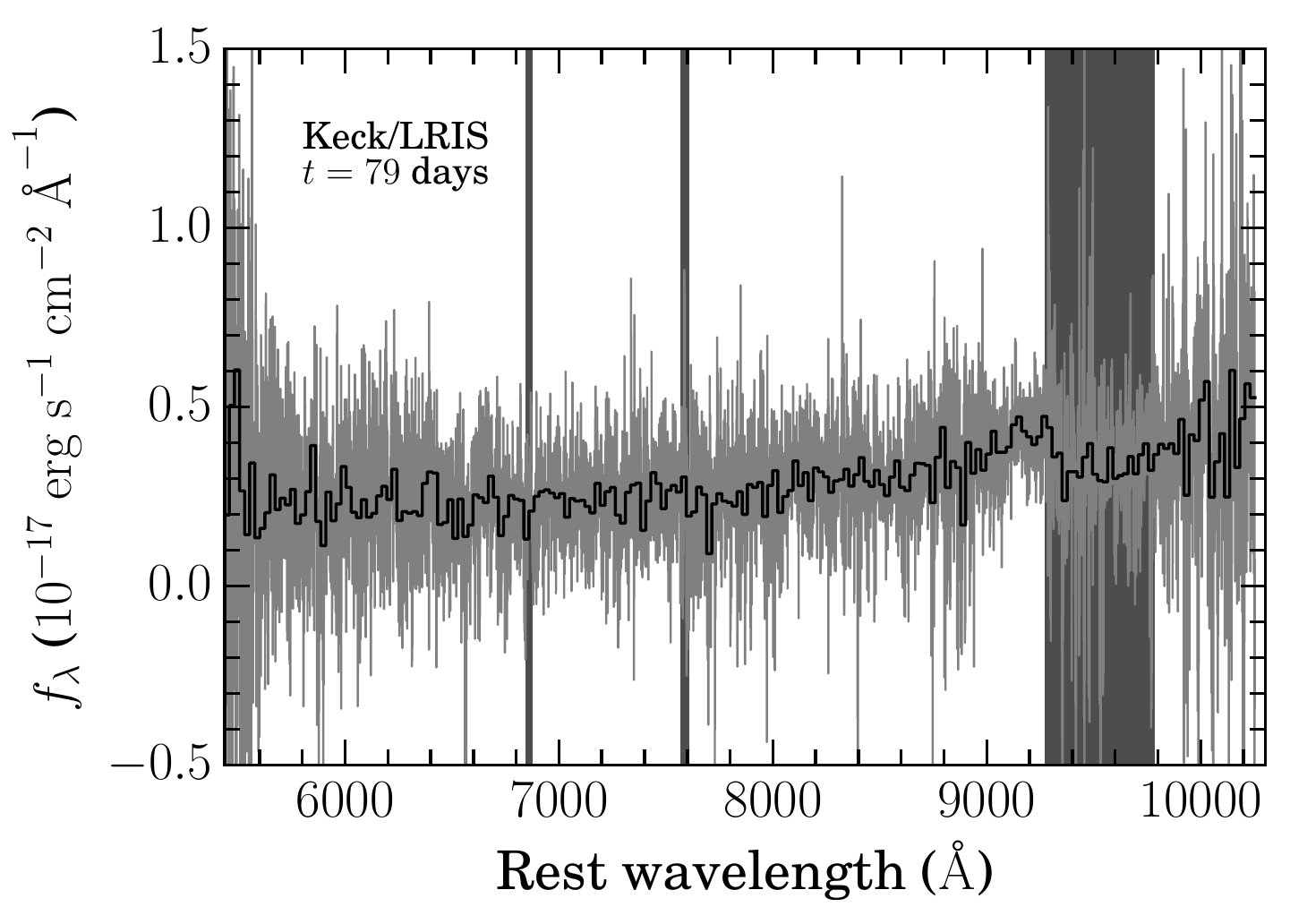}
\caption{\label{fig:spec_opt}
Optical spectrum of SPIRITS\,16tn from Keck/LRIS taken on 2016 November 2 ($t=79$~days) in the rest frame of the host galaxy, NGC~3556 ($z = 0.002332$). The data are shown in light gray, and the median-binned spectrum with a bin size of 20 pixels is overplotted in black. Regions affected by telluric absorption features are indicated by the dark grey vertical bands.
}
\end{figure}

We obtained optical spectroscopy of SPIRITS\,16tn with Keck/LRIS on 2017 November 2 ($t = 79$~days post discovery). We used the D560 dichroic to split the light between the red and blue sides, and used the 400/8500 grating on the red side and 300/3400 grism on the blue side. We obtained one 1800~s integration on the blue side, and two 860~s integrations on the red side. Spectroscopic reductions were performed using the analysis pipeline LPIPE\footnote{Software available at: \url{http://www.astro.caltech.edu/~dperley/programs/lpipe.html}}. A weak trace is visible at the position of the transient on the red-side camera. The low signal-to-noise ratio (S/N) 1D extracted spectrum was flux calibrated using observations of the standard star Feige~34 from the same night. The Keck/LRIS optical spectrum is shown in Figure~\ref{fig:spec_opt}.

We observed SPIRITS\,16tn with the Gemini Near-Infrared Spectrograph (GNIRS) on the 8.1-m Gemini North Telescope on the summit of Mauna Kea in Hawaii through Gemini Fast Turnaround program GN-2016B-FT-25. We obtained two epochs\footnote{Our observations were submitted as a single observation to the Gemini queue, but the execution of our program was split between two separate dates instead, possibly due to deteriorated weather conditions.} of near-IR cross-dispersed (XD; multi-order) spectroscopy on 2016 December 29 ($t=136$~days) and 2017 January 9 ($t=147$~days) using a 0.45~arcsec wide slit with the 32~l/mm grating and the short blue camera with its cross-dispersing prism for a spectra resolution of $R = 1200$. In this configuration, a spectrum of the entire near-IR region (0.85--2.5~$\mu$m) is obtained at once. The observations were carried out using 300~s exposures, with the target nodded along the slit between frames to allow for accurate subtraction of the sky background. We obtained a total of 70~min of integration during the first epoch, and 50~min during the second. Baseline calibrations were also obtained, including observations of A0V stars at similar airmass immediately before/after the science observations as near-IR standards for flux calibration and telluric corrections.

Reductions, including detector pattern noise cleaning, radiation event removal, flat-fielding, background subtraction, spatial distortion corrections, wavelength calibration, and 1D extractions, were perform using standard tasks in the Gemini IRAF\footnote{IRAF is distributed by the National Optical Astronomy Observatory, which is operated by the Association of Universities for Research in Astronomy (AURA) under a cooperative agreement with the National Science Foundation.} package following the procedures outlined on the Gemini webpage\footnote{Procedures for reducing GNIRS XD spectra here:\\\url{http://www.gemini.edu/sciops/instruments/gnirs/data-format-and-reduction/reducing-xd-spectra}}. In the reduced 2D spectra, a faint trace was visible at the position of SPIRITS\,16tn in spectral orders 3 and 4, corresponding to the $K$ and $H$ spectral regions, respectively. 

\begin{figure*}
\begin{minipage}{180mm}
\centering
\includegraphics[width=\linewidth]{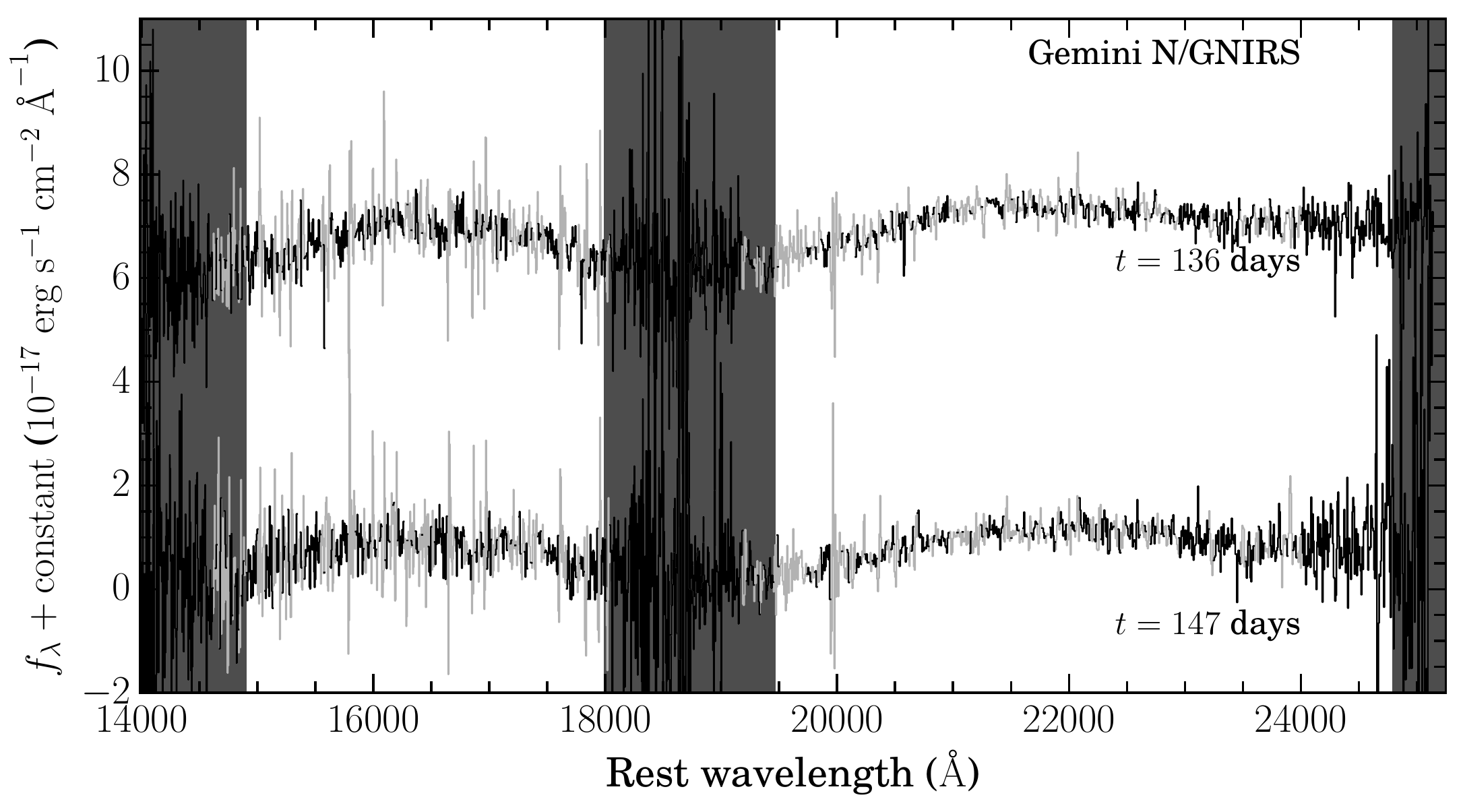}
\caption{\label{fig:spec_nir}
Near-IR spectra of SPIRITS\,16tn from GNIRS on Gemini N. Spectral bins of lower S/N due to coincidence with an OH emission lines of the night sky are shown in light gray. Regions of low atmospheric transmission at the end of the $H$ and $K$ windows are indicated by the grey vertical bars. The spectra have been shifted in wavelength to the rest frame of the host galaxy, NGC~3556 ($z = 0.002332$). The spectrum from 2016 December 29 has been shifted up by the constant indicated on the figure for clarity. 
}
\end{minipage}
\end{figure*}

Corrections for the strong near-IR telluric absorption features and flux calibrations were performed using the IDL tool \textit{xtellcor} developed by \citet{vacca03}. For the first epoch, a detector bias fault occurred during the science target observations, after which the target had to be reacquired. We reduced the two groups of data separately, using the star HIP53735, observed immediately preceding SPIRITS\,16tn, as the A0V flux standard for the first group, and HIP56147, observed immediately after, for the second. The two telluric-corrected, flux calibrated spectra were then averaged. For the second epoch we again used HIP53735. Our Gemini/GNIRS spectra of SPIRITS\,16tn are shown in Figure~\ref{fig:spec_nir}.

We note that we did not attempt to subtract the contribution from the host-galaxy background from our optical/near-IR spectra, which may be significant for our late-time observations as the transient fades. 

\subsubsection{Host spectroscopy}
The Palomar Cosmic Web Imager (PCWI; \citealp{matuszewski10}) is an integral field spectrograph mounted on the Cassegrain focus of the 200-inch Hale telescope at Palomar observatory. The instrument has a field of view of 40 by 60~arcsec divided across 24 slices with dimensions of 40 by 2.5~arcsec each. The spectrograph uses a R $\sim 5000$ volume phase holographic grating (in the red filter) to achieve an instantaneous bandwidth of $\approx 550$ \AA. A complete description of the instrument, observing approach and data analysis methodology can be found in \citealt{martin14}.

We observed the host region of SPIRITS 16tn with PCWI (centered at the location of the transient) on 2017 October 18 in order to characterize the star formation rate in the transient environment. The instrument was configured to a central wavelength of 6630 \AA, covering the wavelength range from approximately 6400 \AA\ to 6900 \AA. We obtained one 600 s exposure of the transient region with the instrument oriented to a position angle of 270$^\circ$ (slices oriented in the North-South direction) followed by one 600 s background sky exposure of a nearby field with no bright sources.

We also obtained calibration images including arc lamp spectra, dome flats and a standard star spectrum. The two dimensional spectra were sliced, rectified, spatially aligned and wavelength calibrated using the calibration images to produce data cubes for each sky exposure, sampled at (RA, Dec., $\lambda$) intervals of (2.6~arcsec, 0.6~arcsec, 0.22 \AA). The sky background cube was subtracted from the source cube to remove the sky emission lines, followed by flux calibration using the standard star Feige 15. This produces the final flux calibrated spectral cube of the 40 by 60~arcsec region centered at the location of the transient.

\subsection{Radio observations}\label{sec:radio_obs}
We observed SPIRITS\,16tn in the radio with the Karl G. Jansky Very Large Array (JVLA) at two epochs on 2016 September 9.0 ($t = 19.0$~days) in the S, C, and X-bands (3, 6, and 10~GHz, respectively) and 2017 January 12.4 ($t = 149.4$~days) in the C and Ku-bands (10 and 15.5~GHz, respectively). The data were reduced using standard imaging techniques for the JVLA in CASA. We also obtained radio imaging at $15$~GHz with the Arcminute Microkelvin Imager Large Array (AMI-LA) on 2017 September 2--5 ($t=17-20$~days). The AMI-LA data were processed (RFI excision and calibration) with a fully-automated pipeline, AMI-REDUCE \citep[e.g.,][]{davies09,perrott13}, and later imported and imaged in CASA. SPIRITS\,16tn was undetected at all epochs and frequencies in the radio, and we provide a summary our limits on the observed fluxes and radio luminosities of the source in Table~\ref{table:radio}. 5-$\sigma$ limits are calculated as 5 times the RMS noise at the position of the transient in the final radio images. 
 
\begin{deluxetable*}{lcccccc}
\tablecaption{Summary of radio observations of SPIRITS\,16tn \label{table:radio}}
\tablehead{\colhead{UT Date} & \colhead{MJD} & \colhead{Phase\tablenotemark{a}} & \colhead{Inst.} & \colhead{Frequency} & \colhead{Flux\tablenotemark{b}} & \colhead{Luminosity\tablenotemark{b,c}}\\ 
\colhead{} & \colhead{} & \colhead{(days)} & \colhead{} & \colhead{(GHz)} & \colhead{(mJy)} & \colhead{(erg~s$^{-1}$~Hz$^{-1}$)} } 
\startdata
2016 Sept 2--5 & 57633 - 57636 & 17--20 & AMI-LA  & 15.0 & $< 0.3$   & $< 2.8 \times 10^{25}$ \\
2016 Sept. 4.0 & 57635.0       & 19.0   & JVLA & 10.0 & $< 0.047$ & $< 4.4 \times 10^{24}$ \\
2016 Sept. 4.0 & 57635.0       & 19.0   & JVLA & 6.0  & $< 0.075$ & $< 7.0 \times 10^{24}$ \\
2016 Sept. 4.0 & 57635.0       & 19.0   & JVLA & 3.0  & $< 0.10$  & $< 9.3 \times 10^{24}$ \\
2017 Jan. 12.4 & 57765.4       & 149.4  & JVLA & 15.5 & $< 0.029$ & $< 2.6 \times 10^{24}$ \\
2017 Jan. 12.4 & 57765.4       & 149.4  & JVLA & 6.0  & $< 0.029$ & $< 2.7 \times 10^{24}$ \\
\enddata
\tablenotetext{a}{Phase is number of days since the earliest detection of this event on 2016 August 15.0 ($\mathrm{MJD} = 57615.0$).}
\tablenotetext{b}{5-$\sigma$ limiting magnitudes are given for non-detections.}
\tablenotetext{c}{Luminosities calculated assuming a distance to NGC~3556 from NED of $8.8$~Mpc.}
\end{deluxetable*}

\section{Analysis} \label{sec:analysis}
Here, we describe our analysis of the both archival imaging data at the position of the transient and our photometric and spectroscopic measurements of SPIRITS\,16tn obtained as part of our follow-up effort. 

\subsection{Progenitor constraints and host environment}\label{sec:progenitor}
We examined the \textit{Spitzer}/IRAC pre-explosion images of NGC~3556 from 2011 February 7, which we also used as references for image subtraction as described in Section~\ref{sec:Spitzer}, for the presence of a possible IR progenitor star. No clear point source is detected in either IRAC channel to 5-$\sigma$ limiting magnitudes of $[3.6] > 14.6$~mag and $[4.5] > 14.4$~mag, where the depth is primarily limited by bright, spatially varying background emission from the host. At the assumed distance to NGC~3556 and correcting only for Galactic extinction, the limits on the absolute magnitudes of the progenitor are $M_{[3.6]} > -15.1$~mag and $M_{[4.5]} > -15.3$~mag.

Images of NGC~3556 were obtained with the \textit{HST} on 1994 July 4 in program SNAP-5446 (PI: G. Illingworth). These observations used the Wide Field Planetary Camera 2 (WFPC2) with the F606W filter, and covered the site of SPIRITS\,16tn, approximately 22 years before its outburst. To determine the precise location of SPIRITS\,16tn in the archival WFPC2 F606W image, we registered this frame with the WFC3 F814W detection image of the active transient described above in Section~\ref{sec:space_imaging}. Using centroid measurements for 10 bright stars detected in both frames, we determined the geometric transformation from WFPC2 to WFC3 using the STSDAS\footnote{STSDAS (Space Telescope Science Data Analysis System) is a product of STScI, which is operated by AURA for NASA.} {\tt geomap} task. By applying the {\tt geotran} task to the WFPC2 frame, and blinking this transformed image against the WFC3 image, we verified the quality of the registration.  The rms error of the geometric fits for the reference stars were 0.15 and 0.20 pixels in the x and y coordinates, respectively, corresponding to $0.006$ and $0.008$~arcsec. We did not detect a source consistent with the position of SPIRITS\,16tn to a 5-$\sigma$ limiting magnitude of $V \gtrsim 24.5$~mag in the archival WFPC2 F606W image. We show the location of SPIRITS\,16tn in the WFPC2 F606W image in the bottom left panel of Figure~\ref{fig:images}. 

At the distance to NGC~3556 and correcting for only Galactic extinction, this corresponds to an upper limit on the absolute magnitude of the progenitor star of $M_V \gtrsim -5.2$. Assuming a bolometric correction for an intermediate red supergiant (RSG) spectral type M0 of $-1.23$~mag from \citet{levesque05}, and adopting a solar bolometric magnitude of $M_{\odot,\mathrm{bol}} = +4.74$~mag, we obtain a limit on the luminosity of the progenitor of $L < 2.9\times10^4~L_{\odot}$. However, if we assume $A_V \approx 7.8$~mag, as inferred to SPIRITS\,16tn below in Section~\ref{sec:sed}, our limit on the progenitor luminosity becomes far less constraining at $L \lesssim 3.9\times10^7~L_{\odot}$.

\begin{figure*}
\begin{minipage}{180mm}
\centering
\includegraphics[width=\linewidth]{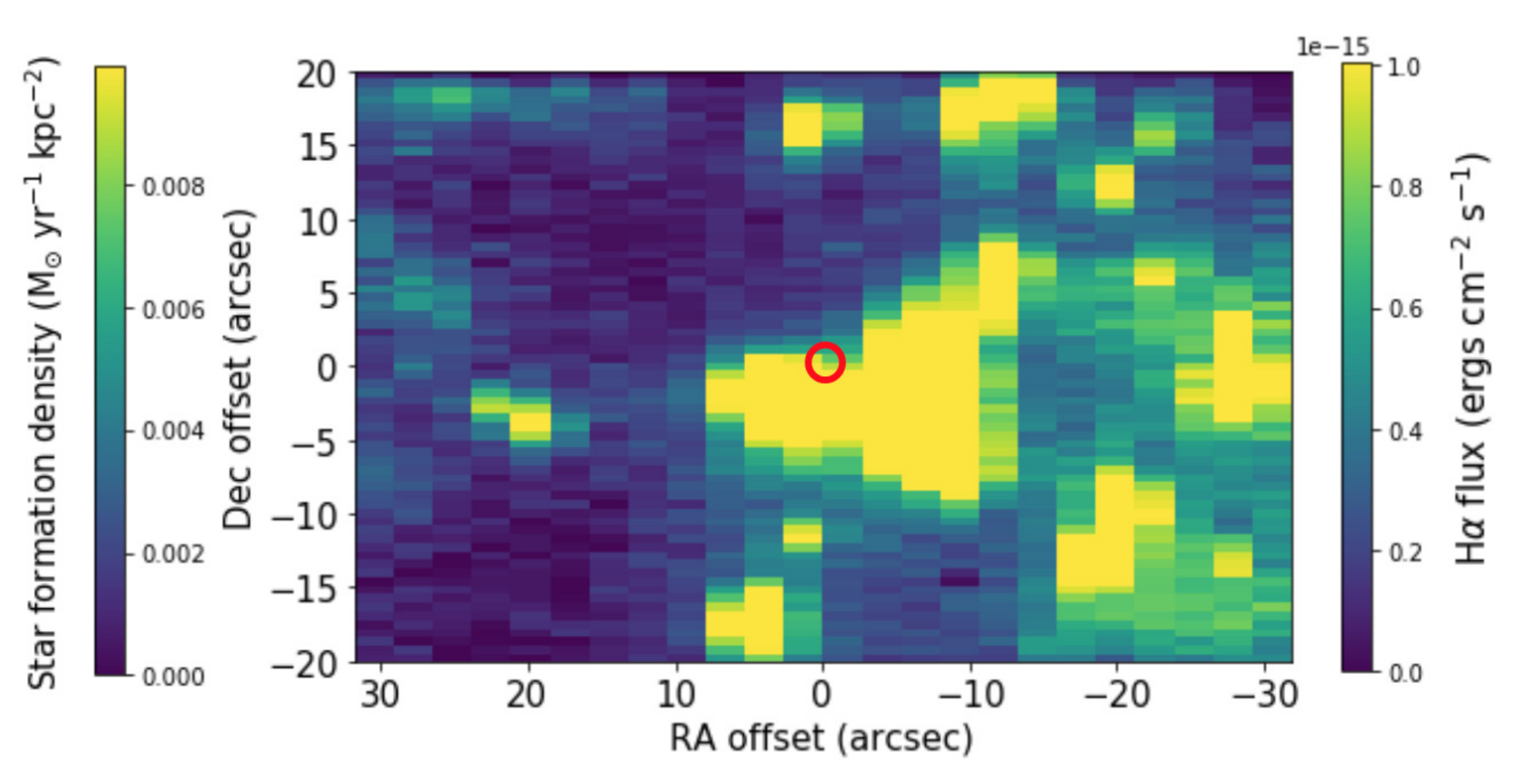}
\caption{\label{fig:pcwi_map}
PCWI map of the host region around SPIRITS\,16tn. The red circle denotes the location of the transient. The image colors correspond to the indicated H$\alpha$ fluxes (right vertical axis) along with their equivalent star formation rate densities (left vertical axis).}
\end{minipage}
\end{figure*}

We use our PCWI observations to constrain the environment of the progenitor of SPIRITS\,16tn. For each pixel in the processed data cube, we fit a simple polynomial to remove the continuum emission from the galaxy. We then measured the H$\alpha$ flux in each pixel by integrating over the H$\alpha$ emission line at the known velocity of the galaxy ($\approx 699$~km~s$^{-1}$), to produce a two dimensional map of H$\alpha$ flux near the location of the transient (Figure~\ref{fig:pcwi_map}). The fluxes were then transformed to an equivalent luminosity using the distance to the galaxy, followed by conversion to an estimated star formation rate using the relations in \citet{kennicutt98}. The star formation rates in each pixel were then converted to an equivalent star formation surface density using an angular scale of 0.052~kpc~arcsec$^{-1}$. As shown, SPIRITS\,16tn is coincident with a dense star forming environment with star formation rate densities of $\sim 10^{-2}$ M$_{\odot}$ yr$^{-1}$ kpc$^{-2}$. 

\subsection{Light curves and color evolution}\label{sec:lcs}
SPIRITS\,16tn was discovered at $[4.5] = 13.04\pm0.05$~mag ($M_{[4.5]} = -16.7$~mag; $\lambda L_{\lambda} = 1.3\times10^{7}~L_{\odot}$), as shown in the light curves in Figure~\ref{fig:lcs}. The high IR luminosity suggests an explosive event, likely an SN. At [4.5] the flux is observed to fade at a rate of $0.018$~mag~day$^{-1}$ between $t=0$ and $184.7$~days. In the near-IR $K_s$-band, SPIRITS\,16tn is observed to fade more slowly at a rate of $0.013$~mag~day$^{-1}$.
These are faster than the expected bolometric decline rates of $0.009$~mag~day$^{-1}$  for light curves powered by the radioactive decay of $^{56}$Co \citep[see, e.g.,][]{gehrz88,gehrz90}. In the $H$- and $J$-bands, the observed decline rates are somewhat slower at $0.009\pm0.004$ and $0.007\pm0.003$~mag~day$^{-1}$, respectively. 

At discovery, the IR color is $[3.6] - [4.5] = 0.676 \pm 0.007$~mag. This corresponds to an effective blackbody temperature of $T_{\mathrm{eff}} \approx 970$~K, possibly indicating that emission from warm dust is a significant contributor to the IR luminosity. At $184.7$~days, SPIRITS\,16tn is observed to have faded more rapidly at [3.6] than [4.5], and evolves to a redder IR color of $[3.6] - [4.5] \gtrsim 1.0$~mag ($T_{\mathrm{eff}} \lesssim 700$~K).

\subsection{SED}\label{sec:sed}
\begin{figure*}
\begin{minipage}{180mm}
\centering
\includegraphics[width=\linewidth]{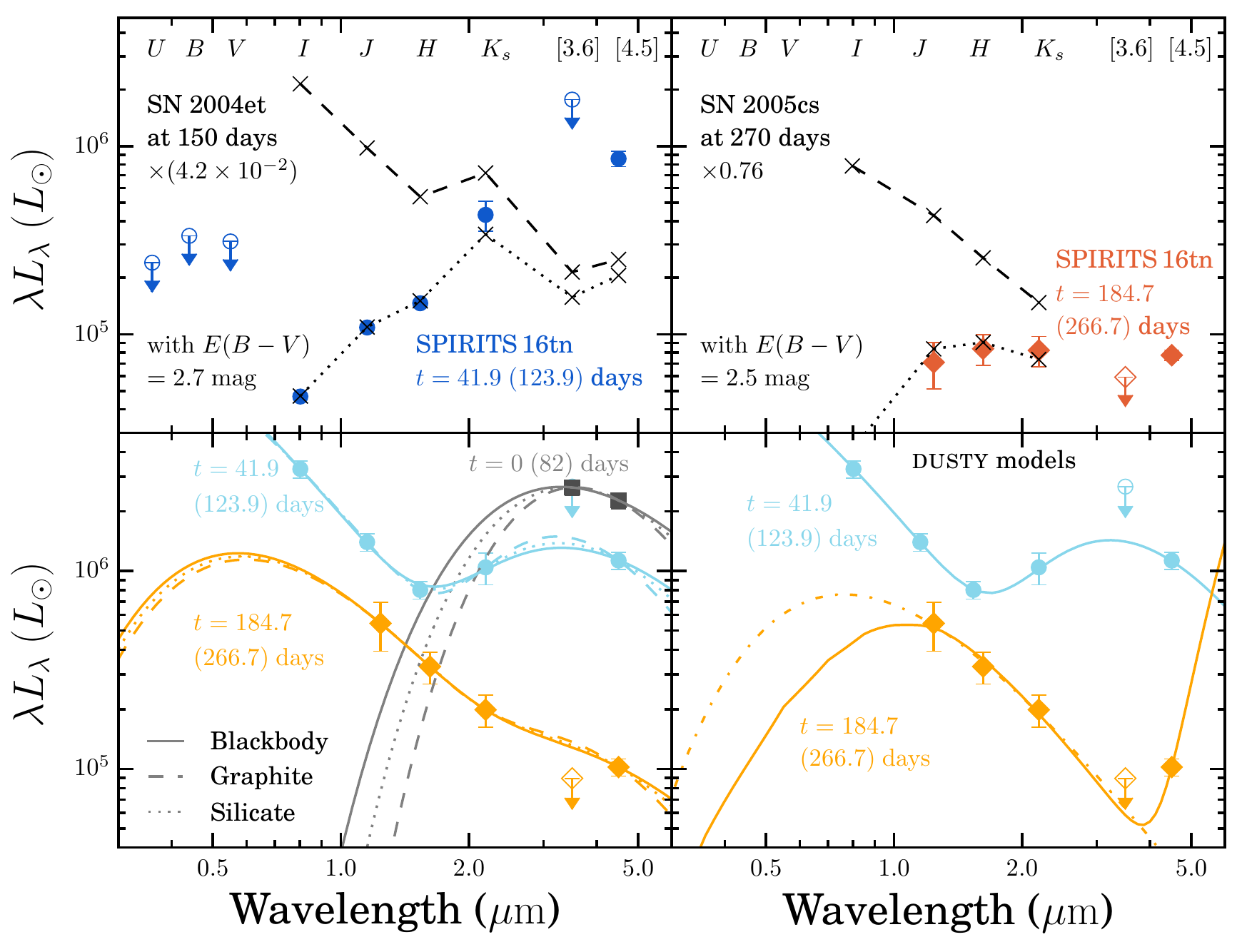}
\caption{\label{fig:sed}
In the top panels, we show the SED evolution of SPIRITS\,16tn, corrected only for Galactic extinction, constructed from the available broadband photometry at nominal phases of $t=41.9$~days (maximum age 123.9~days; dark blue circles) and $t=184.7$~days (maximum age 266.7~days; dark orange diamonds). Unfilled points with downward arrows indicate points treated as upper limits. We compare the observed SED at $t=41.9$~days to that of SN~2004et at a phase of 150~days as the black ``X''-symbols and dashed curve, scaled down in luminosity by a factor of $4.2\times10^{-2}$. The black ``X''-symbols and dotted curve show this SED reddened by $E(B-V) = 2.7$~mag. Similarly, we compare the observed SED of SPIRITS\,16tn at $t=187.4$~days to that of the low-luminosity SN~2005cs at a phase of 270 days scaled down by a factor of 0.76 (dashed black curve) and reddened by $E(B-v) = 2.5$~mag (dashed black curve). In the bottom panels we show the SEDs of SPIRITS\,16tn corrected for an assumed reddening of $E(B-V) = 2.5$~mag at the discovery epoch (gray squares), $t=41.9$~days (light blue circles), and $t=184.7$~days (light orange diamonds). In the bottom left panel, we show multi-component fits to the SEDs using blackbodies (solid curves), and optically thin warm dust components with opacities appropriate for graphite (dashed curves) and silicates (dotted curves). In the bottom right, we show best fits to the SEDs from our dust radiative transfer modeling using \textsc{dusty} as the light blue and orange solid curves. We also show an additional model for the photometry at $t=184.7$~days as the dashed orange curve, treating the [4.5] flux as an upper limit as it is likely enhanced by CO emission at this phase.}
\end{minipage}
\end{figure*}

From our photometry, we constructed quasi-contemporaneous SEDs of SPIRITS\,16tn at two epochs. For the first, we adopt the time of the \textit{HST}/WFC3 detections at $t=41.9$~days as the nominal phase. We use a linear (in magnitudes) interpolation of the [4.5] light curve, and for the $K_s$-band we extrapolate the observed decline back from the detection at $t = 57.5$~days. We include the earlier \textit{Swift}/UVOT non-detections as upper limits. As we only have one detection at [3.6] and cannot interpolate an observed decline rate, we consider this point as an upper limit under the assumption that the transient faded in this band between $t=0$ and $41.9$~days. For the second epoch, we adopt the time of the second [4.5] detection (and non-detection at [3.6]) at $t=184.7$~days as the nominal phase, using extrapolations of $J$ and $H$-band decline rates and interpolating the $K_s$-band light curve. The photometric magnitudes were converted to band-luminosities ($\lambda L_{\lambda}$) at the assumed distance to the host and correcting only for Galactic reddening. To convert the $UBVI$ optical points, we adopt Vega flux zero points and broadband effective wavelengths for the \citet{bessell98} Johnson-Cousins-Glass system. We adopt 2MASS system values from \citet{cohen03} for our $JHK_s$ photometry. For \textit{Spitzer} [3.6] and [4.5] points, we use the flux zero points and effective wavelengths listed in the IRAC instrument handbook. We show the SED evolution of SPIRITS\,16tn in Figure~\ref{fig:sed}.

\subsubsection{Estimating the extinction}\label{sec:SN_ext}
The observed SED at $t=41.9$~days is remarkably red. It is likely that SPIRITS\,16tn suffers from a high degree of host extinction given its location along an obvious dust lane in a highly-inclined, late-type host galaxy. The photometry of SPIRITS\,16tn cannot directly constrain the extinction parameters without some assumptions about the intrinsic SED of the source. We attempt to estimate the extinction to SPIRITS\,16tn by comparing the optical/near-IR SEDs to the SEDs of well-studied SNe.

Type II-Plateau SNe (SNe~IIP) are the most common of all CC SN subtypes, and we use the $IJH$ light curves of the type IIP SN~2004et from \citet{maguire10} as a template for comparison. Following \citet{maguire10}, we adopt an explosion date of UT 2004 September 22.0 ($\mathrm{MJD = 53270.0}$), a distance of $D=5.9$~Mpc, and a total (Galactic and host) extinction parametrized by $E(B-V) = 0.41$~mag for SN~2004et. The light curves of SN~2004et are well-sampled during each of the canonical phases of SN~IIP light curve evolution: the $\approx 100$~day photospheric plateau phase, the rapid fall-off of the plateau, and the subsequent radioactive decline phase. The absolute phase of SPIRITS\,16tn is highly uncertain as our most constraining pre-explosion upper limit is at 82~days before discovery. Using a linear interpolation of the light curves, we find the host extinction parameter $E(B-V)$ which best reproduces the $I-J$ color of SN~2004et between 50 and 150~days in increments of 1~day. Even across this wide range of possible phases, the color evolution of SN~2004et is such that we find a range of $2.6 < E(B-V)~[\mathrm{mag}] < 2.95$, with a mean value of $E(B-V) = 2.8$~mag. 

In the top-left panel of Figure~\ref{fig:sed}, we show the observed SED of SPIRITS\,16tn at $t=41.9$~days (maximum age 123.9~days) compared to the SED of SN~2004et at $150$~days post explosion, just after the start of radioactive decline phase, but scaled down by a factor of $4.2\times10^{-2}$. Applying $E(B-V)=2.7$~mag then provides a good match between the $I$, $J$ and $H$-band measurements to those of SPIRITS\,16tn. 

We performed a similar analysis using the light curves of SN~2005cs in nearby galaxy M51, a prototypical and well-observed low-luminosity SN~IIP explosion \citep{pastorello06,pastorello09}. With a peak bolometric luminosity $\approx 6\times10^{42}$~erg~s$^{-1}$, SN~2005cs was $\sim 10$ times fainter than SN~2004et. We adopt a distance to SN~2005cs of $7.1$~Mpc \citep{takats06}, a total foreground extinction (Milky Way and host) of $E(B-V) = 0.05$~mag \citep{baron07}, and an explosion date of UT 2005 June 27.5 (MJD = 53548.5) as in \citet{pastorello09}. While the $I$-band light curve is well sampled throughout the plateau, fall-off, and decline-tail, the available near-IR $J$ and $H$-band photometry is more limited. We compare the SED of SN~2005cs at a phase of 270~days to the late time SED of SPIRITS\,16tn at $t=184.7$~days (maximum age 266.7~days) in the top-right panel of Figure~\ref{fig:sed}. We find a suitable match with $E(B-V) = 2.5$~mag and scaling the SED of SN~2005cs by a factor of 0.76. 

We find that assuming a SN~II-like SED for SPIRITS\,16tn indicates a high degree of foreground host extinction in the range of $E(B-V) \approx 2.5$--$3.0$~mag ($A_V \approx 7.8$--$9.3$~mag assuming $R_V = 3.1$), regardless of the absolute phase since explosion. This estimate is high compared to interstellar extinction for typical lines of sight in disk galaxies ($A_V \approx 1$ -- $2$~mag~kpc$^{-1}$), but given high inclination of NGC~3556 and the coincidence of SPIRITS\,16tn with a clear dust lane, it is not unreasonable. A direct comparison of the luminosity of SPIRITS\,16tn is not possible in this analysis because of the large uncertainty in absolute phase. The inferred luminosity of SPIRITS\,16tn at $t=184.7$~days is comparable, however, to a late-phase, low-luminosity, SN~2005cs-like explosion, suggesting SPIRITS\,16tn is likely both heavily obscured and intrinsically faint.

\subsubsection{Blackbody and dust component SED models} \label{sec:sed_BBdust}
The bright IR emission associated with SPIRITS\,16tn likely indicates the presence of warm dust. To model the dust emission, we assume an optically thin distribution of dust of total mass $M_{\mathrm{d}}$, composed of spherical grains of radius $a$, radiating thermally at a single, equilibrium temperature $T_{\mathrm{d}}$. This idealized model is described in more detail by, e.g., \citet{hildebrand83}, \citet{dwek85}, and \citet{fox10}.

The expected flux from the warm dust is 

\begin{equation}\label{eq:dust_flux}
F_{\nu} = M_{\mathrm{d}} \frac{\kappa_{\nu}(a) B_{\nu}(T_{\mathrm{d}})}{D^2}
\end{equation}

\noindent where $B_{\nu}(T_{\mathrm{d}})$ is the Planck blackbody function, $D$ is the distance from the source, and $\kappa_{\nu}(a)$ is the dust mass absorption coefficient. In what follows, we assume a grain size of $a=0.1$~$\mu$m, and use broken power law approximations to the dust mass absorption coefficients for dust composed entirely of either graphite or silicate (derived from Mie theory, see Figure~4 of \citealp{fox10}). To account for host extinction, we assume $E(B-V) = 2.5$~mag based on our comparison with SN~2005cs above\footnote{Here, we use a \citep{cardelli89} extinction $R_V = 3.1$ to deredden the photometry to match our analysis using the dust radiative code \textsc{dusty} below in Section~\ref{sec:dusty}.}.

We fit the IR photometric data for the \textit{Spitzer} discovery epoch ($t=0$~days) with this simple dust model for both graphite and silicate compositions to infer $T_{\mathrm{d}}$ and $M_{\mathrm{d}}$. Under the assumption of optically thin dust, the IR luminosity is not sensitive to the size of the dust cloud. In the optically thick case, however, the radius corresponding to blackbody emission provides a lower bound on the dust radius, $R_{\mathrm{d}}$. We repeat the procedure for the quasi-contemporaneous optical--IR SEDs at $t=41.9$ and $187.4$~days, including an additional, hotter blackbody component of temperature $T_*$ and radius $R_*$. We note that for $E(B-V) = 2.5$, our optica/near-IR data do not cover the peak of the hotter SED component, and furthermore, the values inferred for $T_*$ and $R_*$ strongly depend on the choice of extinction. We do not attempt to make strong statements about the properties of the hotter component of the SED for these reasons, and focus our analysis on the properties of the dust component. The results are shown in the bottom-left panel of Figure~\ref{fig:sed} and summarized in Table~\ref{table:SEDmodel}.

\begin{deluxetable*}{ccccccc}
\tablecaption{Results of blackbody and dust component SED modeling \label{table:SEDmodel}}
\tablehead{ \colhead{Phase\tablenotemark{a}} & \colhead{$T_*$} & \colhead{$\log R_*/\mathrm{cm}$} & \colhead{$T_{\mathrm{d}}$} & \colhead{$\log R_{\mathrm{d}}/\mathrm{cm}$} & \colhead{$M_{\mathrm{d}}$} & \colhead{Dust type} \\
\colhead{(days)} & \colhead{(K)} & \colhead{} & \colhead{(K)} & \colhead{} & \colhead{($10^{-4}$~$M_{\odot}$)} & \colhead{} } 
\startdata
0.0   & \nodata & \nodata & 1100 & 15.6    & \nodata & blackbody \\
      & \nodata & \nodata & 680  & \nodata & 1.1     & graphite  \\
      & \nodata & \nodata & 880  & \nodata & 1.5     & silicate  \\
41.9  & 17000   & 13.7   & 1100  & 15.4    & \nodata & blackbody \\
      & 14000   & 13.7   & 730   & \nodata & 0.4     & graphite  \\
      & 15000   & 13.7   & 900   & \nodata & 0.7     & silicate  \\
187.4 & 6400    & 13.9   & 960   & 14.9    & \nodata & blackbody \\
      & 6100    & 13.9   & 660   & \nodata & 0.04    & graphite  \\
      & 6300    & 13.9   & 810   & \nodata & 0.07    & silicate  \\
\enddata
\tablenotetext{a}{Phase is number of days since the earliest detection of this event on 2016 August 15.0 ($\mathrm{MJD} = 57615.0$).}
\end{deluxetable*}

At $t=0$~days, our best fitting results to the IR SED give $T_\mathrm{dust} \approx 680$~K (880~K) and $M_\mathrm{dust} \approx 1.1\times10^{-4}~M_{\odot}$ ($1.5\times10^{-4}~M_{\odot}$) for graphite (silicate) dust. The blackbody fit to the data gives a higher temperature of $T_{\mathrm{d}} \approx 1100$~K and sets a lower bound on the dust radius of $R_{\mathrm{d}} \gtrsim 4.0\times10^{15}$~cm. 

At $t=41.9$~days, we find similar results for the dust temperature for each model, but infer a somewhat lower mass of $M_\mathrm{dust} \approx 0.4\times10^{-4}~M_{\odot}$ ($0.7\times10^{-4}~M_{\odot}$) for graphite (silicate) dust, and a smaller bound on the dust radius from the blackbody fit of $R_{\mathrm{d}} \gtrsim 2.5\times10^{15}$~cm. We note, however, that the uncertainties in interpolating the [4.5] and $K_{s}$-band light curves to construct the SED $t=41.9$~days may have artificially lowered these values. 

At $t=187.4$~days, the data are no longer well fit by a hot source component and warm dust. The best-fitting results find $T_{\mathrm{d}} \approx 700$--$1000$~K, but are notably inconsistent with the upper limit at [3.6]. The blackbody dust radius is smaller by a factor of 3--5 compared to the earlier epochs, and the inferred dust masses are lower by at least a factor of 10. As discussed below in the context of a CC SN, the flux at [4.5] at this phase is likely enhanced by emission from the fundamental vibrational transition of CO and is not attributable solely to thermal emission from warm dust. The lack of evidence for a warm dust component at this phase may indicate the dust has cooled, shifting the flux to longer wavelengths not probed by our data. Alternatively, as discussed below in Section~\ref{sec:echo}, the early presence and subsequent disappearance of the warm dust component may be interpreted as a evidence for an IR echo, i.e., the reprocessing of the UV/optical emission from the luminosity peak of the transient into the thermal IR by a shell of pre-existing circumstellar dust.

\subsubsection{SED modeling with \textsc{dusty}}\label{sec:dusty}
In addition to the simple dust models described above, we modeled the SED of SPIRITS\,16tn using the dust radiative transfer code \textsc{dusty} \citep{ivezic97,ivezic99,elitzur01}. To find best-fitting models and allowed parameter ranges, we use a Markov Chain Monte Carlo (MCMC) wrapper around \textsc{dusty}. Here, we use a spherically symmetric distribution of graphite dust from \citet{draine84} with a standard MRN grain size distribution ($dn/da \propto a^{-3.5}$, $0.005 < a < 0.25$~$\mu$m; \citealp{mathis77}). For the central luminosity source, we assume a simple blackbody spectrum, and allow the model to find best-fitting values for the source temperature, $T_{*}$, and luminosity, $L_{*}$, the optical depth in $V$-band due to circumstellar dust, $\tau_V$, the dust temperature, $T_{\mathrm{d}}$, and inner dust radius, $R_{\mathrm{d}}$. The source temperature and luminosity depend strongly on the assumed foreground extinction, but for simplicity, we again fix the extinction at $E(B-V) = 2.5$~mag as above. Our implementation of \textsc{dusty} uses a \citet{cardelli89} extinction law with $R_V =3.1$, but the differences from the \citet{fitzpatrick99} law assumed throughout the majority of this work are small at the wavelengths of interest in the optical and near-IR.

In this model, the dust is heated by the central source, i.e., the inferred properties of the dust are not independent of $L_*$ and $R_*$. As $T_{\mathrm{d}}$ is constrained strongly by the shape of the IR SED, we do not expect it to vary strongly with the other model parameters. Furthermore, it is fairly robust to the choice of extinction, as the effects of reddening are small in the IR. We expect, however, $R_{\mathrm{d}}$ and $\tau_V$ to vary strongly with the central source properties, namely a hotter, more luminous central source will force dust at a given temperature to larger radii and correspondingly lower optical depths. This model also does not account for light travel time effects inherent to dust at large radii from an evolving, transient source. 

The results of our \textsc{dusty} modeling at both epochs are given in Table~\ref{table:dusty}, including both the values for each parameter for the best-fitting model, i.e., the model that minimizes $\chi^2$, and the median value and 90\% confidence interval limits from the MCMC posterior distributions. We note the the best-fitting values are sometimes near the extrema of the posterior distributions. The best fitting SEDs are also shown in the bottom-right panel of Figure~\ref{fig:sed} in comparison to the observations. 

\begin{deluxetable*}{ccccccccccc}
\tablecaption{Results of SED modeling with \textsc{dusty}\tablenotemark{a} \label{table:dusty}}
\tablehead{ \colhead{Phase\tablenotemark{b}} & \multicolumn{2}{c}{$\log L_*/L_{\odot}$} & \multicolumn{2}{c}{$T_*$} & \multicolumn{2}{c}{$\tau_V$} & \multicolumn{2}{c}{$T_{\mathrm{d}}$} & \multicolumn{2}{c}{$\log R_{\mathrm{d}}/\mathrm{cm}$}\\
\colhead{(days)} & \multicolumn{2}{c}{} & \multicolumn{2}{c}{(K)} & \multicolumn{2}{c}{} & \multicolumn{2}{c}{(K)} & \multicolumn{2}{c}{} } 
\startdata
$41.9$ & $7.4$ & $7.1^{+0.3}_{-0.2}$ & $15200$ & $10900^{+4500}_{-2700}$ & $0.06$ & $0.2^{+0.2}_{-0.1}$ & $840$ & $810^{+80}_{-90}$ & $16.8$ & $16.7^{+0.2}_{-0.2}$ \\
$187.4$ & $8.6$ & $6.2^{+2.0}_{-0.5}$ & $40200$ & $5800^{+24100}_{-2600}$ & $4.9$ & $0.9^{+4.5}_{-0.8}$ & $190$ & $370^{+400}_{-170}$ & $19.1$ & $16.9^{+1.8}_{-1.1}$ \\
$187.4$\tablenotemark{c} & $6.0$ & $6.4^{+1.6}_{-0.7}$ & $48100$ & $7600^{+23800}_{-4100}$ & \nodata & \nodata & \nodata & \nodata & \nodata & \nodata \\
\enddata
\tablenotetext{a}{For each parameter, we give the value for the best-fitting model that minimizes $\chi^2$, and the median value from the MCMC with the 90\% confidence interval limits.}
\tablenotetext{b}{Phase is number of days since the earliest detection of this event on 2016 August 15.0 ($\mathrm{MJD} = 57615.0$).}
\tablenotetext{c}{Results for $t=187.4$~days when [4.5] flux treated as an upper limit including only a single blackbody component.}
\end{deluxetable*}

At $t=41.9$~days, there is a clear IR excess requiring a warm dust component in addition to the interior, hotter source component. The MCMC results for this model have a dust temperature of $T_{\mathrm{d}} = 810^{+80}_{-90}$~K, consistent with the best-fitting value from our simple, optically thin graphite dust model at this epoch in Section~\ref{sec:sed_BBdust}, and an inner dust radius of $R_{\mathrm{d}} = 5.0^{+2.9}_{-1.9}\times10^{16}$~cm at 90\% confidence. As expected, $T_{\mathrm{d}}$ is well constrained by the $K_{s}$-band and [4.5] fluxes and does not vary strongly with the other parameters of the model. While producing a significant IR excess, the dust is optically thin at $\tau_V = 0.2^{+0.2}_{-0.1}$. Because our photometry does not cover the peak of the hot blackbody component, our measurements can only place a lower limit on the temperature and luminosity of the source, and the upper confidence limits found by the MCMC are not physically meaningful. We infer $T_* \gtrsim 8200$~K and $L_* \gtrsim 7.9\times10^6~L_{\odot}$, but note that these limits are highly dependent on our choice of foreground host extinction, i.e., for $E(B-V)_{\mathrm{host}}<2.5$~mag, a lower blackbody temperature and luminosity would be consistent with the data.

At $t=187.4$~days, the hot component has faded by a factor of $\approx 2.7$ in the $J$ and $H$ bands. Again, our photometric measurements do not cover the peak of this component, and thus, the results of the MCMC modeling only allow us to estimate a lower limit on the source temperature of $T_* > 3200$~K. The flux in the redder bands, however, has faded more quickly. The \textsc{dusty} model fits a dust component to the excess flux at [4.5] with constraints from the MCMC at $T_{\mathrm{d}} = 370^{+400}_{-170}$~K, but there is a strong degeneracy between the cooler dust temperatures at smaller radii ($\approx 6\times10^{15}$~cm), and warmer dust at large radii ($\approx 5\times10^{19}$~cm). Again, it is likely that the flux at [4.5] is enhanced by emission from the fundamental vibrational mode of CO, and therefore, is probably not attributable to thermal emission from dust. Treating the measurement at [4.5] instead as an upper limit, we find that the SED of SPIRITS\,16tn at $t=187.4$~days can be adequately modeled with a single blackbody component with $T_* > 4500$~K and $L_* > 5.0\times10^5~L_{\odot}$.

\subsection{The optical and near-IR spectra}
At $t= 79$~days after first detection, the optical spectrum of SPIRITS\,16tn is characterized by a faint, red continuum. There are no clearly discernible features. The apparent dip in the spectrum near $9300$~\AA\ is coincident with a strong telluric absorption band, and is probably not intrinsic to the source. 

The near-IR spectra of SPIRITS\,16tn are shown in Figure~\ref{fig:spec_nir}. Though the Gemini N/GNIRS spectra covered the entire near-IR spectral range from 8500--25000~\AA, we detect emission from SPIRITS\,16tn only in the $H$ and $K$ regions of the spectrum. Due to uncertainty in the age of SPIRITS\,16tn at discovery, the phase of SPIRITS\,16tn is only constrained to be between 136 and 229 days since explosion at the time the spectra were taken.

As in the optical spectrum, we detect a red continuum associated with SPIRITS\,16tn, but there are no unambiguous features in the near-IR. Though the spectra appear to peak near the centers of the $H$ and $K$ spectral windows, we suspect this may an artifact of low S/N and poor flux calibration, particularly near the edges of the bands where little flux is received through the atmosphere. As expected given the high degree of reddening inferred from the SED, we detect the strongest continuum emission in the $K$ spectral region, with an overall decrease in flux toward the blue.

\section{Discussion}\label{sec:discussion}
Here, we compare the observed properties of SPIRITS\,16tn to those of various SNe subtypes to inform our interpretation of the observations. 

\subsection{Comparison to SNe~Ia}
The deep radio non-detections of SPIRITS\,16tn (Section~\ref{sec:radio_obs}) may be easily explained if it is an SN of Type Ia. No SN Ia has been detected as a radio source to deep limits in radio luminosity as far down as $L_{\nu} \lesssim 10^{24}$~erg~s$^{-1}$~Hz$^{-1}$ for the nearest events \citep[e.g.][]{panagia06,chomiuk16}. To test the hypothesis that SPIRITS\,16tn is an SN~Ia, we compare its IR color evolution and near-IR spectrum to well-studied events.

\begin{figure*}
\begin{minipage}{180mm}
\centering
\includegraphics[width=\linewidth]{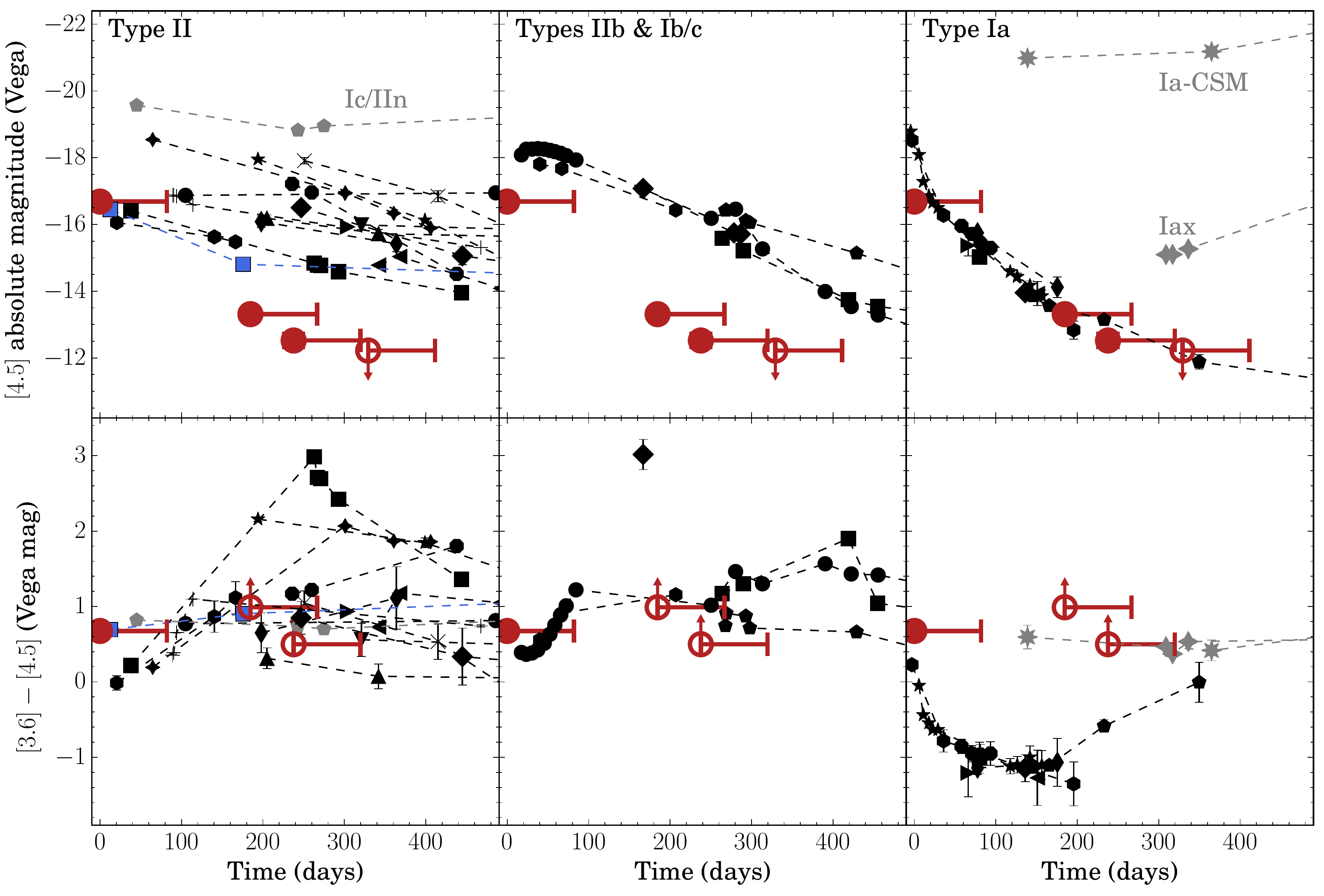}
\caption{\label{fig:SNe_mir}
$[4.5]$ light curves (top row) and $[3.6] - [4.5]$ color evolution (bottom) for SPIRITS\,16tn (red circles) compared to SNe~II (left column), stripped-envelope SNe~IIb and Ib/c (center column), and thermonuclear SN~Ia (right column) shown in black. Lower limits in color are indicated by unfilled points and upward arrows. Time is given on the x-axes as days since discovery for the core-collapse events and days since $B$-band maximum for SNe~Ia. The uncertainty in the phase for SPIRITS\,16tn is indicated by the red horizontal error bars. The sample of SNe~IIP shown in the right column include the \textit{Spitzer}/IRAC measurements for SN~2004A (thick diamonds), SN~2005ad (thin diamonds), SN~2005af (stars), SN~2006my (upward triangles), SN~2006ov (downward triangles), SN~2007oc (``X''-symbols) from \citet{szalai13} and references therein, SN~2011ja (circles), SN~2013am (leftward triangles), SN~2013bu (rightward triangles), SN~2013ej (octagons), and SN~2014bi (squares) from \citet{tinyanont16} and references therein, SN~2004dj (``+''-symbols; \citealp{kotak05,szalai11,meikle11}), SN~2004et (four-point stars; \citealp{kotak09}), and SPIRITS\,14buu (hexagons; \citealp{jencson17}). Also shown is the interaction powered type Ic/IIn SN~2014C \citep[][and references therein]{tinyanont16} as the grey pentagons and the luminous infrared transient SN~2008S \citep{adams16a} as the blue squares. The stripped-envelope SNe shown in the center column include SN~2011dh (circles; type~IIb), SN~2013df (squares; type IIb), SN~2013dk (pentagons; type~Ic), and SN~2014L (hexagons, type~Ic) from \citet{tinyanont16} and references therein, as well as the more recent event SPIRITS\,15C (diamonds; type Ib or IIb, \citealp{jencson17}). Measurements for SNe~Ia from \citet{johansson17} and references therein are shown in the right column for SN~2005df (thin diamonds), SN~2006X (diamonds), SN~2007af (leftward triangles), SN~2007le (rightward triangles), SN~2007sr (squares), SN~2009ig (hexagons), SN~2011fe (pentagons), SN~2012cg (octagons), and SN~2014J (stars). We also show the unusual type Iax SN~2014dt \citep[][and references therein]{fox16} and the interaction powered type Ia-CSM SN~2005gj \citep{fox13b} as gray 4- and 8-pointed stars, respectively. Color measurements for each object are corrected only for Galactic extinction to their respective hosts from NED. Error bars are sometimes smaller than the plotting symbols.}
\end{minipage}
\end{figure*}

In the upper-right panel of Figure~\ref{fig:SNe_mir}, we show the [4.5] light curve of SPIRITS\,16tn compared to several SNe~Ia from \citet{johansson17}, who found that SN~Ia form a homogenous class of objects at these wavelengths. The phases of the SPIRITS\,16tn observations are shown as days since maximum, with $t=0$ assumed to be at the time of the \textit{Spitzer} discovery observations and where the uncertainty in the time of maximum light is indicated by the horizontal error bars. SPIRITS\,16tn shows a similar decline in luminosity at [4.5] to the sample of SNe~Ia, but the $[3.6] - [4.5]$ color evolution, shown in the bottom-right panel of Figure~\ref{fig:SNe_mir}, is notably inconsistent. At discovery, SPIRITS\,16tn had a very red $[3.6] - [4.5]$ color of $0.7$~mag, and evolved to an even redder color of $[3.6] - [4.5] > 1.0$~mag over a period of 185~days. In contrast, SNe~Ia, which may be somewhat red at early times, evolve quickly to the blue reaching $[3.6] - [4.5] \approx -1$~mag at a phase of $\approx 150$~days. SNe~Ia may evolve again to redder colors at very late times, but the observed color of SPIRITS\,16tn is too extreme for SNe~Ia across the entire range of phases relevant here.

Redder $[3.6] - [4.5]$ colors have been observed during the first $400$~days in some thermonuclear SNe, e.g., the interaction powered type Ia-CSM SN~2005gj \citep{fox13b} and the unusual, dusty type~Iax SN~2014dt \citep{fox16}. We show these SNe as the the gray symbols in the upper-right panel of Figure~\ref{fig:SNe_mir}. Both events have an observed late-time, IR flux excess at [4.5] over a normal SN~Ia light curve extending past 200~days, indicative of emission from warm dust. In the context of SNe~Ia light curves, SPIRITS\,16tn does not show such a late-time excess despite its red color.

We have not considered the effects of extinction from the host galaxy or local environment of the SN, e.g., from circumstellar dust. To produce a color excess of $0.7$~mag between [3.6] and [4.5] would require an additional $A_V \gtrsim 40$~mag of extinction, using the empirically derived broadband extinction parameters for [3.6] and [4.5] from \citet{chapman09}, much higher than the inferred extinction to SPIRITS\,16tn from the SED of $A_V \approx 8$~mag.

\begin{figure}
\centering
\includegraphics[width=\linewidth]{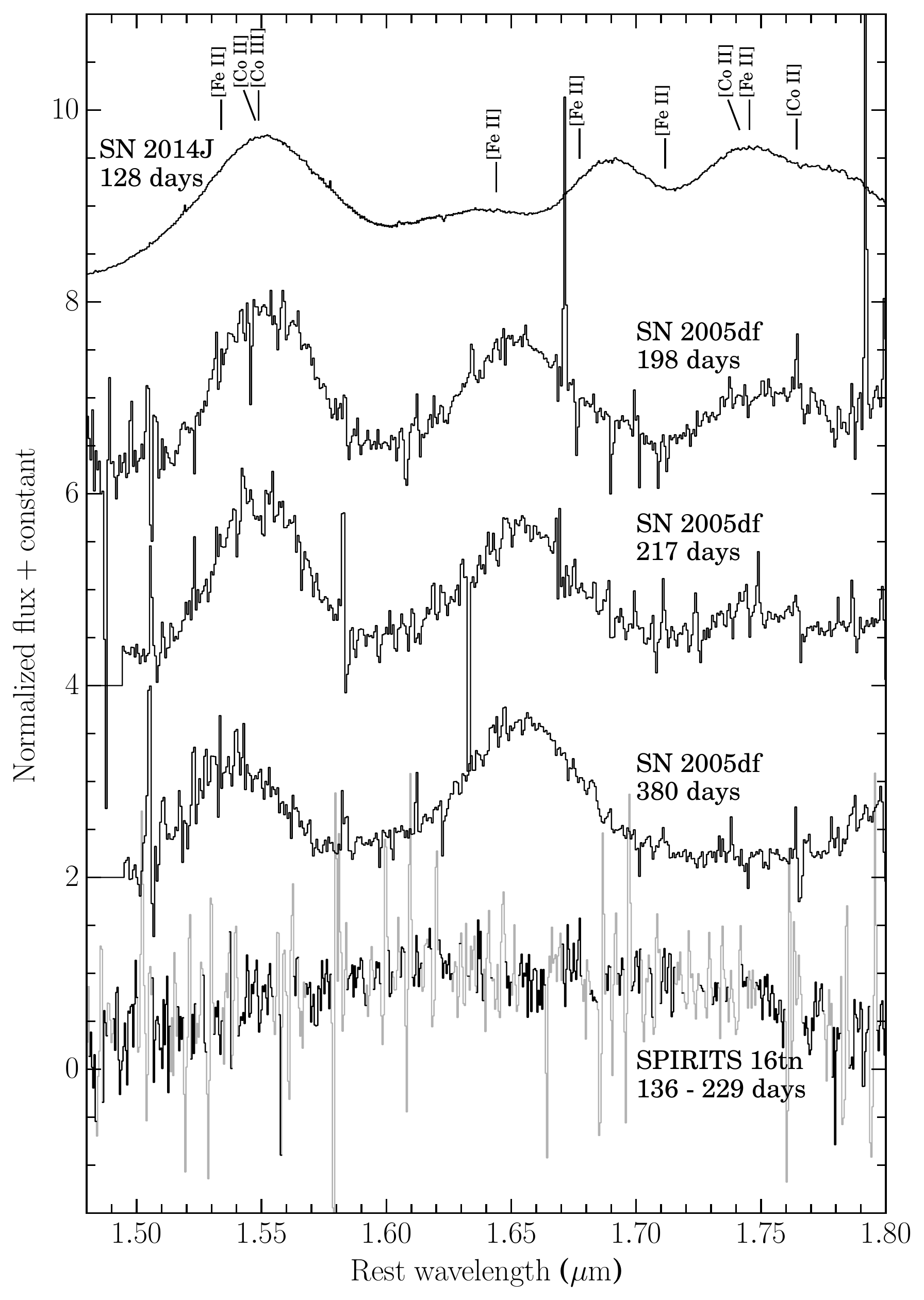}
\caption{\label{fig:spec_Ia}
$H$-band spectrum of SPIRITS\,16tn, at a phase between 136--229~days, compared to late-time, nebular spectra the type Ia SN~2014J \citep{johansson17} and SN~2005df \citep{diamond15}. The SPIRITS\,16tn spectrum is the average of the two Gemini N/GNIRS spectra taken on 2016 December 29 and 2017 January 9. Spectral bins of lower S/N due to coincidence with an OH emission line of the night sky are shown in light gray. The spectra have been normalized by the flux at $1.644~\mu$m, and each spectrum is shifted up from the one below for clarity. Forbidden transitions of Fe-peak elements identified in \citet{diamond15} are indicated above the spectrum of SN~2014J. These broad, blended features are not present in SPIRITS\,16tn.}
\end{figure}

Furthermore, the featureless spectrum of SPIRITS\,16tn is wholly inconsistent with normal SNe~Ia at comparable phases. In Figure~\ref{fig:spec_Ia}, we show the $1.48 - 1.9~\mu$m region ($\approx H$-band) of the spectrum of SPIRITS\,16tn, along with the late-time spectra of type~Ia SN~2014J at a phase of 128 days (post $B$-band maximum) from \citet{johansson17}, and the type~Ia SN~2005df at 198, 217, and 380 days \citep{diamond15}. The phase of our near-IR spectrum is constrained to be between 136--229 days. The late-time spectra of SNe~Ia are dominated by blended, nebular emission features, primarily forbidden transitions of Fe-peak elements \citep{bowers97,spyromilio04}. In the $H$-band region, \citet{diamond15} specifically identified transitions of [Fe~\textsc{ii}], [Co~\textsc{ii}], and [Co~\textsc{iii}] in the spectra of type~Ia SN~2005df, as labeled in Figure~\ref{fig:spec_Ia}. They noted that the spectrum became Fe-dominated as the Co features faded between $\approx 200$--$380$~days, however the strong, broad emission lines of [Fe~\textsc{ii}] persisted to very late phases. These features are completely absent from the near-IR spectrum of SPIRITS\,16tn, and thus, we definitively rule out a reddened SN~Ia in this case.

\subsection{Comparison to CC SNe}\label{sec:ccsne}
We compare the light curve at [4.5] of SPIRITS\,16tn to CC SNe of type~II and stripped-envelope types~IIb and Ib/c the upper left and center panels of Figure~\ref{fig:SNe_mir}. Among the hydrogen-rich type II SNe, we do not distinguish between the photometric subtypes IIP and IIL, defined by the presence of a light curve plateau or linear decline, respectively. Our dataset for SPIRITS\,16tn is insufficient to make such a distinction. Furthermore, the existence of the two truly distinct subclasses is debated and recent studies of large SN~II samples have suggested that type IIP and IIL SNe may instead form a continuous distribution in their observed properties \citep[e.g.,][]{anderson14,sanders15,rubin16}. 

The observed [4.5] luminosity peak of SPIRITS\,16tn at $−16.7$~mag is in the range of type~II SNe, but is $\gtrsim 1$~mag fainter than is observed for the sample of stripped-envelope events. Notably, SPIRITS\,16tn fades more rapidly at [4.5] at $0.017$~mag~day$^{-1}$ than any of the CC SNe for which comparable data were available. The fastest event in the comparison sample is the type II SN~2013ej (black octagons in the upper-left panel of Figure~\ref{fig:SNe_mir}) fading at a rate of $0.013$~mag~day$^{-1}$. Although the sample of stripped-envelope SNe is small, they appear relatively more homogeneous at [4.5] compared to SNe~II, with typical decline rates between $0.009$--$0.012$~mag~day$^{-1}$. Given the larger degree of variation in both peak luminosity at [4.5] and the observed decline rate for SNe~II, it is easier to reconcile the lower [4.5] peak and faster decline of SPIRITS\,16tn with the sample of SNe~II. 

As shown in the bottom panels of Figure~\ref{fig:SNe_mir}, SPIRITS\,16tn develops a very red color by $t=184.7$~days of $[3.6] - [4.5] > 1.0$~mag. Similarly red colors have been observed at comparable phases for several CC SNe including the type~IIP events SN~2005af \citep[][and reference therein]{szalai13}, SN~2004et \citep{kotak09}, and SN~2014bi \citep{tinyanont16}, and for the type~IIb/Ib event SPIRITS\,15C \citep{jencson17}. While a red mid-IR color may be a signature of thermal emission from warm dust ($T_{\mathrm{eff}} \lesssim 700$~K for $[3.6] - [4.5] > 1.0$~mag), emission from the 1--0 vibrational transition of CO at $\approx 4.65~\mu$m can produce excess flux at [4.5] compared to the other mid-IR bands. This emission feature has been directly identified in the mid-IR spectra of several type~II SNe including SN~1987A \citep[e.g.,][]{meikle89,wooden93}, SN~2004dj \citep{kotak05}, and SN~2005af \citep{kotak06}. Corroborating the identification of this feature, the bandheads of the $\Delta \nu = 2$ vibrational overtones of CO, which produce excess emission beyond 2.3~$\mu$m at the end of the $K$-band, have also been observed in, e.g., SN~1987A \citep{meikle89}, SN~2004dj \citep{kotak05}, and the stripped-envelope events SN~2011dh \citep{ergon15} and SPIRITS\,15C \citep{jencson17}. We do not clearly detect this feature in our near-IR spectra from $t = 136$--$145$~days, but note that the $\Delta \nu = 2$ vibrational overtones may be significantly weaker than the fundamental band at [4.5] and hidden in our low S/N spectra. Furthermore, the spectra were obtained at an earlier epoch, possibly before CO formed in the ejecta. In the context of CC SNe, we consider CO emission to be the most likely explanation for the observed mid-IR color evolution of SPIRITS\,16tn, indicating the presence of CO in the ejecta by $t \approx 185$~days.

In Figure~\ref{fig:lcs}, we compare the multi-band light curves of SPIRITS\,16tn to those of the low-luminosity type IIP SN~2005cs. The light curves of SN~2005cs are shifted to the distance of SPIRITS\,16tn and reddened with $E(B-V) = 2.5$~mag, as inferred from our SED comparison in Section~\ref{sec:sed}. For the relative phase offset shown, and with an additional offset in apparent magnitude of $\Delta m = 0.7$~mag (factor of $\approx 2$ in flux), the late-time $IJHK_{s}$ light curves can be reasonably well-matched to those of SPIRITS\,16tn. In this scenario, our \text{HST} observations of SPIRITS\,16tn at $t=41.9$~days would have occurred just after the transition to the nebular phase and require a plateau duration $\lesssim 123.9$~days to be consistent with our $z$-band pre-explosion non-detection. Given our lack of early-time date for SPIRITS\,16tn and the notable gap in near-IR photometric coverage of SN~2005cs during the transition to the nebular phase, we cannot perform a more detailed light curve comparison. Still, we find the optical--near-IR light curve evolution of SPIRITS\,16tn to be largely consistent with a SN~2005cs-like, low-luminosity type~IIP event.

\begin{figure}
\centering
\includegraphics[width=\linewidth]{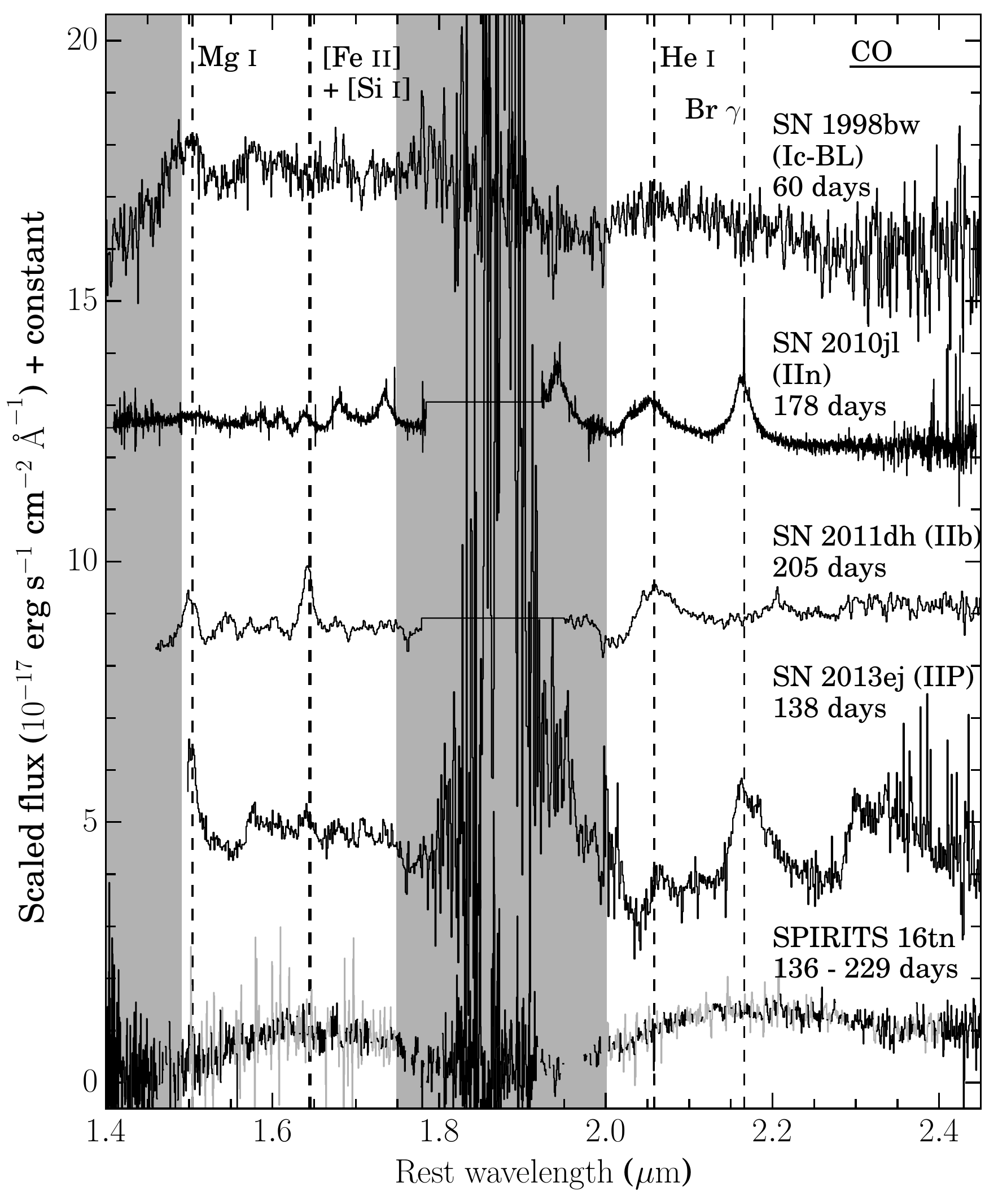}
\caption{\label{fig:spec_CC}
We show the $H$ and $K$-band spectra of SPIRITS\,16tn at a phase between 136--229~days along with the late-phase spectra of CC SNe of various types for comparison. The CC SNe spectra are scaled in flux to the distance of SPIRITS\,16tn and reddened by $E(B-V) = 2.5$~mag. Each one is shifted up from the one below by an arbitrary constant for clarity. Prominent features present in the spectra of some CC SNe are indicated by the dashed vertical lines and labeled near the top of the figure.}
\end{figure}

In Figure~\ref{fig:spec_CC}, we compare near-IR spectrum of SPIRITS\,16tn at a phase between 136--229~days post maximum to those of CC SNe of various types including the type~Ic broad-lined (Ic-BL) SN~1998bw \citep{patat01}, the type IIn SN~2010jl \citep{borish15}, the type IIb SN~2011dh \citep{ergon15}, and the type II SN~2013ej \citep{yuan16}. Some of the most prominent features identified in late-phase CC SNe in the $H$ and $K$-bands are labeled in Figure~\ref{fig:spec_CC}, including Mg~\textsc{i} at $1.504~\mu$m, blended [Fe~\textsc{ii}] at $1.644~\mu$m and [Si~\textsc{i}] at $1.646~\mu$m, \ion{He}{1} at $2.058~\mu$m, Br~$\gamma$ at $2.166~\mu$m, and the bandheads of the $\Delta \nu = 2$ vibrational overtones of CO beyond $2.3~\mu$m. Pa~$\alpha$, a typically strong H~\textsc{i} feature in SNe~II, is unfortunately in the low atmospheric transmission region between the $H$ and $K$ spectral windows where we did not receive any detectable flux from SPIRITS\,16tn. 

As we do not detect any clear features in SPIRITS\,16tn, we are unable to provide a definitive classification. However, the lack of clear a spectroscopic signature of the interaction of the SN ejecta with a dense CSM, often observed as superimposed narrow ($\sim \mathrm{few} \times 100$~km~s$^{-1}$) and broad ($\sim \mathrm{few} \times 1000$~km~s$^{-1}$) components of the H~\textsc{i} and He~\textsc{i} features, can rule out a strongly interacting SN~IIn. We suggest that at late times, it is possible the near-IR spectral features of non-interacting CC SNe may be very weak.

\subsubsection{Radio limits}\label{sec:radio_SNe}
\begin{figure}
\centering
\includegraphics[width=\linewidth]{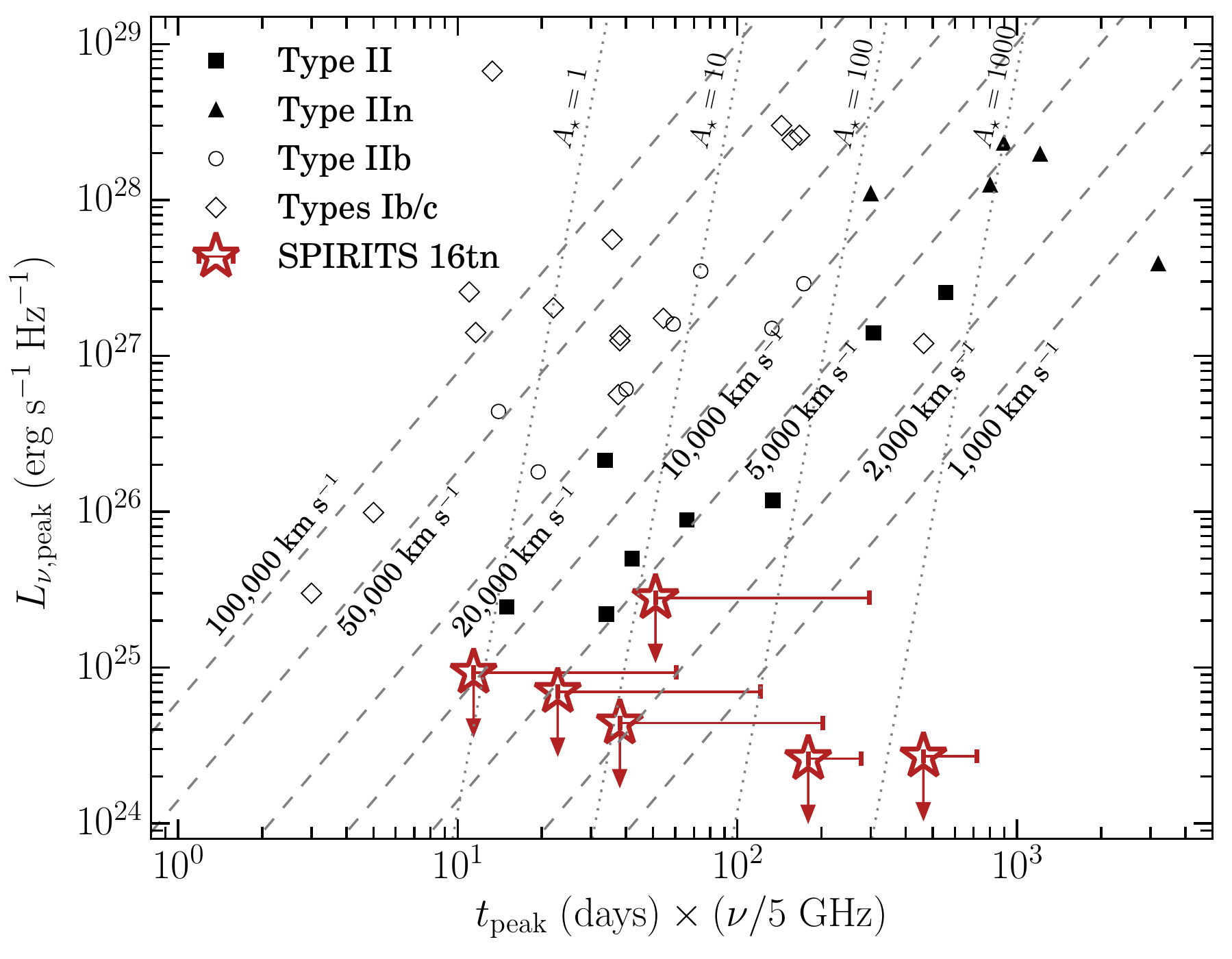}
\caption{\label{fig:radio}
Peak radio luminosity vs.\ time of peak times the frequency of observation for radio CC SNe adapted from \citep[e.g.][]{chevalier06a,romero-canizales14}. Type IIs are shown as black squares, and interaction-powered SNe~IIn are shown as black triangles. Unfilled circles and diamonds represent stripped-envelope SNe~IIb and Ib/c, respectively. Upper limits on the radio luminosity of SPIRITS\,16tn at a given phase from our non-detections are shown as red, unfilled stars with downward arrows, where the horizontal errorbars represent our uncertainty in the absolute phase since explosion. Assuming a SSA model with an electron distribution with $p=3$ for the shockwave propagating through the CSM, one can infer the shock velocity (dashed lines) and CSM density parameter ($A_{\star}$; dotted lines) from the position on this diagram.}
\end{figure}

In Figure~\ref{fig:radio}, we show our limits on the radio luminosity of SPIRITS\,16tn as a function of phase compared to the peak radio luminosities and times to peak for CC SNe. Radio emission is produced in CC SNe when the fastest SN ejecta interact with and shock the slow-moving pre-explosion CSM from the pre-explosion stellar wind of the progenitor. As the shockwave propagates through the CSM, turbulent instabilities amplify magnetic fields and accelerate relativistic electrons \citep{chevalier82}. The resultant radio emission is characterized by slowly declining, optically thin, non-thermal synchrotron and early, optically thick absorption at low frequencies. Proposed absorption mechanisms include synchrotron self-absorption (SSA) or internal free-free absorption in the emitting region and free-free absorption by the external, ionized CSM \citep[e.g.,][]{chevalier82,chevalier98}.

If one assumes SSA is dominant, for an electron population with an energy spectral index of $p=3$, the size of the radio emitting region at the time of the SSA peak can be calculated as \citep{chevalier98}

\begin{equation}\label{eq:Rs}
\begin{split}
	R_{\mathrm{s}} = 4.0 \times 10^{14} \alpha^{-1/19} \left(\frac{f}{0.5} \right)^{-1/19} \left(\frac{F_{\mathrm{p}}}{\mathrm{mJy}} \right)^{9/19} \\
	     \times \left(\frac{D}{\mathrm{Mpc}}\right)^{18/19} \left(\frac{\nu}{5~\mathrm{GHz}}\right)^{-1} \mathrm{cm},
\end{split}
\end{equation}

\noindent where $\alpha \equiv \epsilon_e/\epsilon_B$ is the ratio of the energy density in relativistic electrons to that in the magnetic field, $f$ is the fraction of the spherical volume filled by the radio emitting region, $F_p$ is the peak flux at frequency $\nu$, and $D$ is the distance to the source. If additional absorption mechanisms are important, this radius must be even larger. The shockwave velocities, $v_{\mathrm{s}}$, inferred for $\alpha = 1$ (assuming equipartition) and $f = 0.5$ (as estimated in \citealp{chevalier06b}) are shown as the dashed lines in Figure~\ref{fig:radio}. We the note weak dependence in Eq.~\ref{eq:Rs} of the shock radius on these parameters. 

For a steady, pre-SN stellar wind, the density profile of the CSM as a function of radius, $r$, is $\rho_{\mathrm{w}} = A/r^2 \equiv \dot M / (4 \pi r^2 v_{\mathrm{w}})$, where $\dot M$ is the mass loss rate and $v_{\mathrm{w}}$ is the wind velocity. $A \equiv \dot M / (4 \pi v_{\mathrm{w}})$ is the normalization of the CSM density profile as in \citet{chevalier82}. As calculated by \citet{chevalier06b}, the radio emission at time $t$ since explosion of a CC SN is sensitive to the density profile of the CSM as

\begin{equation}\label{eq:Astar}
\begin{split}
	A_{\star} \epsilon_{B-1} \alpha^{8/19} = 1.0 \left(\frac{f}{0.5} \right)^{-8/19} \left(\frac{F_{\mathrm{p}}}{\mathrm{mJy}} \right)^{-4/19} \\
	     \times \left(\frac{D}{\mathrm{Mpc}}\right)^{-8/19} \left(\frac{\nu}{5~\mathrm{GHz}}\right)^{2} \left(\frac{t}{10~\mathrm{days}}\right)^{2},
\end{split}
\end{equation}

\noindent where $\epsilon_{B-1} \equiv \epsilon_B/0.1$ and $A_{\star} \equiv A/(5 \times 10^{11}~\mathrm{g}~\mathrm{cm}^{-1})$ is a dimensionless proxy for $A$. We show lines of constant $A_{\star}$ in Figure~\ref{fig:radio}, determined largely by the strong dependence of this parameter on $t_{\mathrm{peak}}$.  

Our deep non-detections of SPIRITS\,16tn indicate that this event is either an intrinsically weak radio source, or that the emission is heavily absorbed. Though we do not rule out radio emission arising at very late times, characteristic of strongly interacting SNe~IIn with a dense CSM, this scenario is unlikely given the lack of prominent interaction features in the optical/near-IR spectra. Our observations are inconsistent with most varieties of stripped-envelope events, which tend to be more luminous radio sources. A high velocity ($v_{\mathrm{s}} \gtrsim 50,000$~km~s$^{-1}$) SN~Ic with a fast-evolving radio light curve, however, is not explicitly ruled out (cf. SN~2007gr, \citealp{soderberg10}, and PTF12gzk, \citealp{horesh13a}). Our optical--IR SED analysis and comparisons with well-studied SNe~II indicate that SPIRITS\,16tn falls at the low end of the SN luminosity function, and a weak SN~II radio counterpart is consistent with our observations. Using Eq.~\ref{eq:Astar}, for typical SN~II shock velocities of $v_{\mathrm{s}} \approx \mathrm{few} \times 10^3$~km~s$^{-1}$, our non-detections can constrain $A_{\star} \epsilon_{B-1} \alpha^{8/19} \lesssim 24$. For a steady pre-SN wind, we then infer a limit on the pre-SN mass loss rate of $\dot M \lesssim 2.4 \times 10^{-6} \left(\frac{\epsilon_{B}}{0.1}\right)^{-1} \alpha^{-8/19} \left(\frac{v_\mathrm{w}}{10~\mathrm{km~s}^{-1}}\right)~M_{\odot}$~yr$^{-1}$. 

Such mass loss rate is consistent with a low-luminosity RSG ($L \approx 10^{4.5}$ -- $10^5 L_{\odot}$) based on the standard observational prescriptions of \citet{dejager88} (See, e.g., Figure~3 of \citealp{smith14}). This supports the picture of SPIRITS\,16tn as a low-luminosity SN~II arising from the explosion of a low-mass ($M \approx 10$ -- $15 M_{\odot}$) RSG progenitor. We note that, depending on the assumed explosion date of SPIRITS\,16tn, for $v_{\mathrm{s}} = 10000$~km~s$^{-1}$ and $v_{\mathrm{w}} = 10$~km~s$^{-1}$, the timing of our radio observations probes the mass loss history of the progenitor only in the final $\approx 50$ -- $600$~yr before the explosion. 

Concurrent panchromatic observations spanning the radio to the X-ray have indicated deviations from energy equipartition in some SNe. For example, for SN~2011dh \citep{soderberg12} find $\alpha=30$ and $\epsilon_{\mathrm{B}} = 0.01$. Adopting such values results in only a modest decrease in the shock velocity (using Eq.~\ref{eq:Rs}) by a factor of 1.2, an an increase in our limit on the mass loss rate by a factor of 2.4. Alternatively for SN~2011dh, \citep{horesh13b} find $\alpha \approx 500$--$1700$, adopting a value of 1000 as a reasonable average, and $\epsilon_{\mathrm{B}} = 3\times10^{-4}$. This still results in only a modest decrease in the inferred shock velocity by a factor of 1.4, and for a fixed wind velocity, the limit on the mass loss rate increases by a factor of 20. 

\subsubsection{Origin of the observed dust component}\label{sec:echo}
In Section~\ref{sec:sed}, our modeling of the SED of SPIRITS\,16tn at $t=0$ and $41.9$~days suggests the presence of an IR component ($T_{\mathrm{d}} \approx 700$--$900$~K) powered by thermal emission by at least $M_{\mathrm{d}} \approx 1.0$--$1.5 \times 10^{-4}~M_{\odot}$ of warm dust (our observations are not sensitive to any additional dust at cooler temperatures). By $t=187.4$~days, we no longer see evidence for this warm dust component, indicating it has either faded or the dust has cooled, shifting the flux to longer wavelengths.

The dust emission may arise either from pre-existing circumstellar dust formed in the pre-SN wind of the progenitor, or newly formed dust in the dense, rapidly cooling ejecta behind the SN shock. For a shock velocity of $v_{s} = 10000$~km~s$^{-1}$, the radius of the shock at a phase of 82.0 days (the maximum age of SPIRITS\,16tn at $t=0$~days) is $R_s = 7.1 \times 10^{15} \left( \frac{v_s}{10^4~\mathrm{km~s}^{-1}} \right)$~cm. As we infer a lower limit on the dust radius (blackbody radius from Section~\ref{sec:sed_BBdust}) of $R_{\mathrm{d}} \gtrsim 4.0\times 10^{15}$~cm that is smaller than that shock radius, it is plausible that the emitting dust is located in the post-shock cooling zone. Given the low observed dust temperatures, however, it is unlikely that the early dust component is due to newly formed dust in the ejecta. We would expect newly formed dust to be near the evaporation temperature, as, given sufficiently high densities, dust grains will begin to condense as soon as the drop in temperature of the radiation environment allows. Typical values for astrophysical dust are $T_{\mathrm{evap}} \approx 1900$~K for graphite, and $T_{\mathrm{evap}} \approx 1500$~K for silicate, significantly hotter than the observed dust component.

For pre-existing dust, we can interpret the observed IR-excess as an IR echo. In this scenario, a pre-existing shell of dust is heated by the peak luminosity of the explosion and re-radiates this energy thermally in the IR. The duration of the echo is related to the size of the dust shell from geometrical arguments as $\Delta t \sim 2 R_{\mathrm{d}}/c$. The observations of the warm dust component between $t = 0$ to $41.9$~days, and subsequent fading by $t=187.4$~days, when the maximum age of SPIRITS\,16tn is $269.4$~days, would then require $5.4\times10^{16} \lesssim R_{\mathrm{d}} \lesssim 3.5\times10^{17}$~cm.

As a consistency check, we can estimate the peak luminosity, $L_{\mathrm{peak}}$, of the transient required to heat spherical dust grains of radius $a$ within this range of distances to the observed temperatures. The energy absorbed by a dust grain is balanced by the energy it radiates as

\begin{equation}\label{ref:abs_em}
\frac{L_{\mathrm{peak}}}{4 \pi R_{\mathrm{d}}^2} \pi a^2 Q_{\mathrm{abs}} = 4 \pi a^2 \sigma_{\mathrm{SB}} T_{\mathrm{d}}^4 Q_{\mathrm{em}}.
\end{equation}

\noindent The peak luminosity of the transient is then given by 

\begin{equation}\label{eq:Lpeak}
L_{\mathrm{peak}} = 16 \pi R_{\mathrm{d}}^2 \sigma_{\mathrm{SB}} T_{\mathrm{d}}^4 \frac{Q_{\mathrm{em}}}{Q_{\mathrm{abs}}},
\end{equation}

\noindent where $Q_\mathrm{em}$ and $Q_{\mathrm{abs}}$ are Planck-averaged emission and absorption efficiencies for \citet{laor93} dust grains. For the temperature of the incident radiation field we assume values $T_{\mathrm{rad}} = 10000$~K and $T_{\mathrm{rad}} = 6000$~K, characteristic of an SN at peak. We use a value of $\sigma_{\mathrm{SB}} = 5.67 \times 10^{-5}$~erg~cm$^{-2}$~K$^{-4}$~s$^{-1}$ for the Stefan-Boltzmann constant. For graphite grains of size $a=0.1~\mu$m at $T_{\mathrm{d}} = 680$~K, we find $4.7\times 10^{40} \lesssim L_{\mathrm{peak}} \lesssim 1.9\times10^{42}$~erg~s$^{-1}$ for $T_{\mathrm{rad}} = 10000$~K, and similar values of $5.7\times 10^{40} \lesssim L_{\mathrm{peak}} \lesssim 2.3\times10^{42}$~erg~s$^{-1}$ for $T_{\mathrm{rad}} = 6000$~K. Alternatively, for grains of silicate composition at $T_{\mathrm{d}} = 880$~K, we find a somewhat higher values of $9.9\times 10^{41} \lesssim L_{\mathrm{peak}} \lesssim 4.0\times10^{43}$~erg~s$^{-1}$ for $T_{\mathrm{rad}} = 10000$~K, or $2.6\times 10^{41} \lesssim L_{\mathrm{peak}} \lesssim 1.0\times10^{43}$~erg~s$^{-1}$ for $T_{\mathrm{rad}} = 6000$~K.

While these estimates are crude, we can still compare them to the observed range of peak luminosities for SNe~II. \citet{faran18}, for example, estimate the bolometric luminosities of 29 SNe~II and find peak values spanning at least two orders of magnitude and ranging from $L_{\mathrm{bol}} \approx 2.4\times10^{41}$--$2.4\times10^{43}$~erg~s$^{-1}$. Similarly, the pseudo-bolometric (from $\sim U$ to $I$-band) light curves of the sample of SN~II from \citet{valenti16} span $L_{\mathrm{bol}} \approx 1.0\times10^{41}$--$5\times10^{42}$~erg~s$^{-1}$. Among the faintest SNe~II known, the quasi-bolometric $UBVRI$ light curve of SN~1999br peaked at only $4.5\times10^{40}$~erg~s$^{-1}$ \citep{pastorello04,pastorello09}. The observed range of peak luminosities for SNe~II is similar to the range of luminosities estimated above to explain the IR-excess of SPIRITS\,16tn as a dust echo. 

We can also estimate the pre-SN mass loss rate of the progenitor star necessary to support the such an echo. We assume the dust is concentrated in a thin shell with $\Delta r/R_{\mathrm{d}} = 0.1$, a dust-to-gas ratio of $M_{\mathrm{d}}/M_{\mathrm{g}} = 0.01$, and again a pre-SN wind velocity of $v_{\mathrm{w}} = 10$~km~s$^{-1}$ and find $\dot M \approx 9\times10^{-6}$ -- $6\times10^{-5} \left(\frac{M_{\mathrm{d}}}{10^{-4} M_{\odot}}\right) \left(\frac{M_{\mathrm{d}}/M_{\mathrm{g}}}{0.01}\right)^{-1} \left(\frac{\Delta r/R_{\mathrm{d}}}{0.1}\right)^{-1}
\left(\frac{v_{\mathrm{w}}}{10~\mathrm{km~s}^{-1}}\right) M_{\odot}$~yr$^{-1}$ for the range of $R_{\mathrm{d}}$ allowed by the observations. These estimates are a factor of $\approx 4$ -- $25$ higher than that in Section~\ref{sec:radio_SNe} based on the radio observations, but probe an earlier time in the mass loss history of the progenitor of somewhere between $\approx 1700$ -- $11000$~yr before explosion. While such mass loss rates would imply a more luminous and massive RSG progenitor ($L \approx 10^{5.2}$ -- $10^{5.5} L_{\odot}$, again assuming standard \citealp{dejager88} prescriptions), if the dust is confined to a shell, this could indicate a relatively brief episode of enhanced mass loss.

\subsection{Non-SN transient scenarios}

As we are unable to provide a definitive classification of SPIRITS\,16tn as a CC SN, we briefly consider other observed classes luminous IR transients as possible explanations for this event.

\subsubsection{SN~2008S and NGC~300 OT2008-1-like transients}
SN~2008S and the luminous 2008 optical transient in NGC~300 (NGC~300 OT2008-1) are the prototypes of a distinct class of transients. They have optically obscured progenitors, but bright mid-IR pre-explosion counterparts ($M_{[4.5]} < -10$~mag), suggested to be extreme asymptotic giant branch stars of intermediate mass ($\approx 10$--$15~M_{\odot}$) self-obscured by a dense, dusty wind \citep{prieto08,bond09,thompson09}. They are less luminous than typical CC SNe at peak ($L_{\mathrm{bol}} \approx 10^{41}$~erg~s$^{-1}$ for SN~2008S, \citealp{botticella09}). Emission lines in their spectra indicate slow expansion velocities of 70--80~km~s$^{-1}$ \citep[e.g.,][]{bond09,humphreys11}. A proposed physical scenario is a weak explosion, possibly an electron-capture SN, or massive stellar eruption that destroys most of the obscuring dust, allowing the transient to be optically luminous. The development of a late-time IR excess, however, suggests the dust reforms, obscuring the optical transient at late times \citep[e.g.,][]{thompson09,kochanek11,szczygiel12}. Both events are now fainter than their progenitor luminosities as [3.6] and [4.5], suggesting the transients were terminal events \citep{adams16a}.  

The [4.5] light curve of SN~2008S is shown as the blue squares in Figure~\ref{fig:SNe_mir} compared to SPIRITS\,16tn and several SNe~II. The peak luminosity at [4.5] is similar to SPIRITS\,16tn. Furthermore, the peak bolometric luminosity of SN~2008S-like events is sufficient to power the IR dust echo discussed in Section~\ref{sec:echo}. The IR luminosity of SPIRITS\,16tn declines more rapidly, and we do not observe a late-time IR excess powered by newly formed dust, inconsistent with the characteristic evolution of SN~2008S-like events.

\subsubsection{Massive stellar mergers}
The 2011 transient in NGC~4490 (hereafter NGC~4490-OT, \citealp{smith16}) and M101 2015OT-1 \citep{blagorodnova17} are proposed massive analogs of the galactic contact binary merger V1309~Sco \citep{tylenda11}, and the B-type stellar merger V838~Mon \citep{bond03,sparks08}. These events typically have unobscured, optical progenitors, irregular, multi-peaked light curves, increasingly red colors with time, and significant late-time IR excesses powered by copious dust formation. Their spectra show relatively narrow emission features of H~\textsc{i} indicating low velocities of $\sim 100$~km~s$^{-1}$. 

In the IR, NGC~4490-OT peaked at $M_{[4.5]} \approx -15$~mag, $\approx 1.7$~mag fainter than SPIRITS\,16tn. The inferred mass for the progenitor system of NGC~4490-OT was 20--30~$M_{\odot}$, and thus, a merger origin for the more luminous SPIRITS\,16tn would likely require an exceptionally massive progenitor. Furthermore, the IR light curve of NGC~4490-OT is long-lived, remaining too luminous at phases $\gtrsim 800$~days to be powered by an IR echo \citep{smith16}. The relatively short-lived IR excess of SPIRITS\,16tn, interpreted here as an IR echo, is generally inconsistent with the observed IR evolution typical of massive star mergers.

\section{Summary and Conclusions} \label{sec:conclusions}
SPIRITS\,16tn is a luminous ($M_{[4.5]} = -16.7$~mag) mid-IR transient discovered with \textit{Spitzer}/IRAC during the ongoing SPIRITS survey in the nearby galaxy NGC~3556. We believe SPIRITS\,16tn is a possible SN. Despite being one of nearest SNe discovered in 2016 at only 8.8~Mpc, it was completely missed by optical searches due to heavy extinction. The transient position is coincident with a dark dust lane in the inclined, star-forming disk of the host. We estimate a total extinction of $A_V = 7.8$--$9.3$~mag, making SPIRITS\,16tn one of the most highly obscured SNe yet discovered in the IR. 

The [4.5] light curve shows a fast decline of $0.018$~mag~day$^{-1}$, and the source becomes increasingly red in the mid-IR from $[3.6]-[4.5] = 0.7$~mag to $\gtrsim 1.0$~mag between $t=0$ and $184.7$~days post discovery. The optical and near-IR spectra display a featureless, red continuum, ruling out an SN~Ia, but preclude a definitive spectroscopic classification. The SED at $t=41.9$~days post discovery is extremely red, and can be matched to an SN~II-like SED with $E(B-V) = 2.5$ -- $3.0$~mag of extinction. Furthermore, our analysis suggest SPIRITS\,16tn may be an intrinsically dim event similar to the well-studied, low-luminosity SN~2005cs. Modeling of the SED indicates the presence of a warm dust component ($T \approx 700$ -- $900$~K), which fades by $t = 184.7$~days. This is consistent with an IR echo powered by a circumstellar shell of dust located somewhere between $5.4\times10^{16}$ -- $3.5\times10^{17}$~cm heated by a peak luminosity of $\sim 5 \times 10^{40}$ -- $4 \times 10^{43}$~erg~s$^{-1}$, similar to the range of observed peak luminosities for SNe~II.

The source is not detected to deep limits in the radio across frequencies of 3--15.5~GHz, constraining the radio luminosity to $\lesssim 4\times10^{24}$~erg~s$^{-1}$~Hz$^{-1}$ between $t=19$ and $149.4$~days. This effectively rules out most stripped-envelope SNe, except possibly the most rapidly evolving, high velocity events that may peak in the radio at very early times. A late-rising, interaction-powered SN~IIn may be consistent with our radio limits, but the typically strong spectroscopic signatures of interaction with a dense CSM are absent from our optical/near-IR spectra. SNe~II, typically the weakest radio emitters among CC SNe, are the most consistent with our deep radio limits, and in this context we can constrain the pre-SN mass loss rate of the progenitor to $\dot M \lesssim 2.4 \times 10^{-6} \left(\frac{\epsilon_{B}}{0.1}\right)^{-1} \left(\frac{v_\mathrm{w}}{10~\mathrm{km~s}^{-1}}\right)~M_{\odot}$~yr$^{-1}$. This is consistent with a lower mass RSG progenitor of $M \sim 10$--$15 M_{\odot}$.

We analyzed the available pre-explosion \textit{Spitzer}/IRAC 2011 imaging, and \textit{HST}/WFPC2 F606W imaging of NGC~3556 covering the site of SPIRITS\,16tn from 1994, and do not detect a candidate progenitor star. Given the high degree of extinction inferred to SPIRITS\,16tn, however, we are unable to place meaningful limits on the progenitor luminosity.

Taken together, we find the most likely explanation for the observed properties of SPIRITS\,16tn to be an SN~II explosion that is both highly obscured by foreground, host-galaxy dust, and intrinsically low-luminosity. This discovery strengthens the fact that, even in the local 10 Mpc volume, SN searches appear to be incomplete. Transient surveys in the IR have the unique ability to find dust obscured or otherwise optically dim events, allowing for the true nearby SN population to be uncovered. 

\acknowledgments
We thank R. Lunnan for executing our Keck/LRIS observations. We thank D. Neill and M. Matuszewski for assistance with the PCWI observations and data reduction. We thank A. Monson for help with P200/WIRC data reduction. We thank T. Cantwell, Y. Perrott, and the AMI staff for conducting the AMI-LA observations and data reduction. 

This material is based upon work supported by the National Science Foundation Graduate Research Fellowship under Grant No. DGE-1144469. H.E.B acknowledges support for this work provided by NASA through grants GO-13935 and GO-14258 from the Space Telescope Science Institute, which is operated by AURA, Inc., under NASA contract NAS 5-26555. R.D.G. was supported in part by the United States Air Force. A.H. acknowledges support by the I-Core Program of the Planning and Budgeting Committee and the Israel Science Foundation.

This work is based in part on observations made with the Spitzer Space Telescope, which is operated by the Jet Propulsion Laboratory, California Institute of Technology under a contract with NASA. The work is based, in part, on observations made with the Nordic Optical Telescope, operated by the Nordic Optical Telescope Scientific Association at the Observatorio del Roque de los Muchachos, La Palma, Spain, of the Instituto de Astrofisica de Canarias.

This work is based in part on observations with the NASA/ESA {\it Hubble Space Telescope\/} obtained at the Space Telescope Science Institute, and from the Mikulski Archive for Space Telescopes at STScI, which are operated by the Association of Universities for Research in Astronomy, Inc., under NASA contract NAS5-26555. These observations are associated with programs GO-14258 and SNAP-5446.

Some of the data presented herein were obtained at the W. M. Keck Observatory, which is operated as a scientific partnership among the California Institute of Technology, the University of California and the National Aeronautics and Space Administration. The Observatory was made possible by the generous financial support of the W. M. Keck Foundation. The authors wish to recognize and acknowledge the very significant cultural role and reverence that the summit of Maunakea has always had within the indigenous Hawaiian community.  We are most fortunate to have the opportunity to conduct observations from this mountain.

RATIR is a collaboration between the University
of California, the Universidad Nacional Autono\'{m}a de
Me\'{x}ico, NASA Goddard Space Flight Center, and Arizona
State University, benefiting from the loan of an H2RG detector from Teledyne Scientific and Imaging. RATIR, the
automation of the Harold L. Johnson Telescope of the
Observatorio Astrono\'{m}ico Nacional on Sierra San Pedro
Ma\'{r}tir, and the operation of both are funded by the partner institutions and through NASA grants NNX09AH71G,
NNX09AT02G, NNX10AI27G, and NNX12AE66G, CONACyT
grant INFR-2009-01-122785, UNAM PAPIIT grant
IN113810, and a UC MEXUS-CONACyT grant. 

UKIRT is owned by the University of Hawaii (UH) and operated by the UH Institute for Astronomy; operations are enabled through the cooperation of the East Asian Observatory. When the data reported here were acquired, UKIRT was supported by NASA and operated under an agreement among the University of Hawaii, the University of Arizona, and Lockheed Martin Advanced Technology Center; operations were enabled through the cooperation of the East Asian Observatory.

Based on observations obtained at the Gemini Observatory acquired through the Gemini Observatory Archive and processed using the Gemini IRAF package, which is operated by the Association of Universities for Research in Astronomy, Inc., under a cooperative agreement with the NSF on behalf of the Gemini partnership: the National Science Foundation (United States), the National Research Council (Canada), CONICYT (Chile), Ministerio de Ciencia, Tecnolog\'{i}a e Innovaci\'{o}n Productiva (Argentina), and Minist\'{e}rio da Ci\^{e}ncia, Tecnologia e Inova\c{c}\~{a}o (Brazil).

The National Radio Astronomy Observatory is a facility of the National Science Foundation operated under cooperative agreement by Associated Universities, Inc.

The Legacy Surveys consist of three individual and complementary projects: the Dark Energy Camera Legacy Survey (DECaLS; NOAO Proposal ID \# 2014B-0404; PIs: David Schlegel and Arjun Dey), the Beijing-Arizona Sky Survey (BASS; NOAO Proposal ID \# 2015A-0801; PIs: Zhou Xu and Xiaohui Fan), and the Mayall z-band Legacy Survey (MzLS; NOAO Proposal ID \# 2016A-0453; PI: Arjun Dey). DECaLS, BASS and MzLS together include data obtained, respectively, at the Blanco telescope, Cerro Tololo Inter-American Observatory, National Optical Astronomy Observatory (NOAO); the Bok telescope, Steward Observatory, University of Arizona; and the Mayall telescope, Kitt Peak National Observatory, NOAO. The Legacy Surveys project is honored to be permitted to conduct astronomical research on Iolkam Du'ag (Kitt Peak), a mountain with particular significance to the Tohono O'odham Nation.

NOAO is operated by the Association of Universities for Research in Astronomy (AURA) under a cooperative agreement with the National Science Foundation.

The Legacy Survey team makes use of data products from the Near-Earth Object Wide-field Infrared Survey Explorer (NEOWISE), which is a project of the Jet Propulsion Laboratory/California Institute of Technology. NEOWISE is funded by the National Aeronautics and Space Administration.

The Legacy Surveys imaging of the DESI footprint is supported by the Director, Office of Science, Office of High Energy Physics of the U.S. Department of Energy under Contract No. DE-AC02-05CH1123, by the National Energy Research Scientific Computing Center, a DOE Office of Science User Facility under the same contract; and by the U.S. National Science Foundation, Division of Astronomical Sciences under Contract No. AST-0950945 to NOAO.

Funding for SDSS-III has been provided by the Alfred P. Sloan Foundation, the Participating Institutions, the National Science Foundation, and the U.S. Department of Energy Office of Science. The SDSS-III web site is http://www.sdss3.org/.

SDSS-III is managed by the Astrophysical Research Consortium for the Participating Institutions of the SDSS-III Collaboration including the University of Arizona, the Brazilian Participation Group, Brookhaven National Laboratory, Carnegie Mellon University, University of Florida, the French Participation Group, the German Participation Group, Harvard University, the Instituto de Astrofisica de Canarias, the Michigan State/Notre Dame/JINA Participation Group, Johns Hopkins University, Lawrence Berkeley National Laboratory, Max Planck Institute for Astrophysics, Max Planck Institute for Extraterrestrial Physics, New Mexico State University, New York University, Ohio State University, Pennsylvania State University, University of Portsmouth, Princeton University, the Spanish Participation Group, University of Tokyo, University of Utah, Vanderbilt University, University of Virginia, University of Washington, and Yale University. 

%\facility{facility ID}
\facilities{Spitzer (IRAC), HST (WFPC2, WFC3), Mayall (CCD Mosaic imager), Swift (UVOT), Hale (WIRC), PO:1.5m (SEDM), Keck:I (LRIS), OANSPM:HJT (RATIR), UKIRT (WFCAM), Gemini:Gillett (GNIRS), EVLA, AMI} 
%\software{Numpy}

\bibliographystyle{yahapj}
\bibliography{SPIRITS16tn}

%\appendix
%\section{appendix section}

\end{document}